\documentclass[aps,prd,twocolumn,amsmath,amssymb,,groupedaddress,nofootinbib,numerical,superscriptaddress]{revtex4-1}

\usepackage{amsmath}
\usepackage{graphicx}
\usepackage{lipsum}
\usepackage{natbib}

\usepackage{multirow}
\usepackage{array,mathtools,amssymb,booktabs}
\newcolumntype{C}{>{$}c<{$}}
\AtBeginDocument{
\heavyrulewidth=.08em
\lightrulewidth=.05em
\cmidrulewidth=.03em
\belowrulesep=.65ex
\belowbottomsep=0pt
\aboverulesep=.4ex
\abovetopsep=0pt
\cmidrulesep=\doublerulesep
\cmidrulekern=.5em
\defaultaddspace=.5em
}
\newcommand{\ra}[1]{\renewcommand{\arraystretch}{#1}}

\renewcommand{\a}{\hat{a}}
\renewcommand{\b}{\hat{b}}

\newcommand{\adot}{\dot{a}}
\newcommand{\addot}{\ddot{a}}

\begin{document}

\title{Non-linear cosmological spherical collapse of quintessence}

\author{J. Rekier${}^*$}
\affiliation{Namur Centre for Complex Systems (naXys), University of Namur, B-5000 Namur, Belgium}
\email[]{jeremy.rekier@unamur.be}
\author{A.F\"uzfa}
\affiliation{Namur Centre for Complex Systems (naXys), University of Namur, B-5000 Namur, Belgium}
\author{I. Cordero-Carri\'on}
\affiliation{Departamento de Matem\'atica Aplicada, Universidad de Valencia, E-46100 Burjassot, Spain}
%\thanks{}

\date{\today}

\begin{abstract}

We present a study of the fully relativistic spherical collapse in presence of quintessence using on Numerical Relativity, following the method proposed by the authors in a previous article~\cite{Rekier2015_1}. We ascertain the validity of the method by studying the evolution of a spherically symmetric quintessence inhomogeneity on a de Sitter background and we find that it has an impact on the local expansion around the centre of coordinates. We then proceed to compare the results of our method to those of the more largely adopted \emph{top-hat} model. We find that quintessence inhomogeneities do build up under the effect that matter inhomogeneities have on the local space-time yet remain very small due to the presence of momentum transfer from the over-dense to the background regions. We expect that these might have an even more important role in modified theories of gravitation.

\end{abstract}

%\pacs{}

\maketitle

\section{Introduction}
\label{section:Introduction}
In a companion paper~\cite{Rekier2015_1}, we proposed a method based on techniques of Numerical Relativity to compute the collapse of a matter over-density on the expanding space-time background. In the present article, this method is extended in order to incorporate the effects of quintessence modelled as a real scalar field. 

The late-time acceleration of the expansion of the Universe is validated by many independent observations including \emph{type-Ia supernovae diagramme}~\cite{Riess:1998cb,Perlmutter:1998np,Riess:2001gk,Kowalski:2008ez,Amanullah:2010vv}, angular fluctuations of the \emph{Cosmic Microwave Background} (CMB)~\cite{Komatsu:2008hk,Ade:2015rim} and \emph{galaxy redshift distortion}~\cite{Guzzo:2008ac}. One can explain this acceleration by \emph{fine-tuning} the value of a cosmological constant $\Lambda$. There is, however, a number of conceptual problems~\cite{Weinberg:1988cp} associated with this that has lead many to consider other options. Among these is the hypothesis that late-time acceleration is due to the dynamics of a new variety of energy called \emph{quintessence}~\cite{Copeland:2006wr,Ratra:1987rm,Wetterich:1987fm}. In this framework, the study of the formation of large scale structure performed with N-Body simulations is a very interesting topic as it allows to draw important conclusions on the dynamics of the homogeneous Universe and the nature of the quintessence component~\cite{Deus2012,Bouillot2015,2005Nature}. All of these studies rely on simplified models of collapse of local inhomogeneities.

Some groups have investigated the collapse in presence of quintessence using the top-hat formalism\footnote{By ``top-hat'', we understand a certain type of model in which the total energy density profile remains a step function throughout the whole evolution. The whole space-time consists only into two parts at all time and not only initially.}~\cite{Wang1998,Voit2004,Horellou2005,Abramo2009,Fernandes2012,Li2014}. Some investigated the case where quintessence is due to a real scalar field~\cite{Mota2004,Nunes2004,Wintergerst2010}. However, the top-hat model can only be somewhat motivated as a limit case of the LTB solution valid for pressure-less matter only and, in spite of its successes, is not a rigorous solution of Einstein equations in the general case. 

In most works, it is argued that quintessence should not cluster on astrophysical scales and contributes only to the background dynamics. The method that we have developed allows to test this hypothesis. It also allows to investigate the evolution and effects of the anisotropic stress due to the presence of quintessence on the formation of structure beyond the perturbation level which constitutes a good test on the nature of quintessence~\cite{Bertschinger:2008zb,Chang:2014bea}. The fully relativistic non-perturbative approach proposed uses tools from Numerical Relativity applied to the evolution of cosmological space-times with a scalar field following the early work by Shibata et al.~\cite{Shibata1999} as well as more recent works in the field~\cite{Torres2014,Alcubierre:2015ipa}. We highlight the role played by the momentum transfer associated with the quintessence field.

The paper is organised as follows. In section~\ref{section:top-hat}, we review the top-hat model with quintessence. The formalism that we used to solve for the fully relativistic solution is adapted from our previous work~\cite{Rekier2015_1} and is detailed in sections~\ref{sec:Formalism} and~\ref{sec:FormalismSF}. The validity of this is first tested in the case of a de Sitter Universe filled with inhomogeneities quintessence only. This is the object of section~\ref{sec:compTH} and~\ref{sec:beyond} respectively concerned with a comparison with the top-hat model and a more detailed study that goes beyond the results from top-hat. Conclusions and perspectives are discussed in section~\ref{section:conclusion}.

\section{The top-hat model}
\label{section:top-hat}

In the top-hat model, space-time and the matter distribution are kept piecewise homogeneous. The Universe is divided into an inner and an outer regions each following its own Friedmann equation~\cite{padmanabhan1993structure,Wang1998,Weinberg2003}~:
\begin{align}
\left(\frac{\dot{a}}{a}\right)^2&=\frac{8\pi}{3}\bar{\rho}~,\\
\left(\frac{\dot{R}}{R}\right)^2&=\frac{8\pi}{3}\bar{\rho}(1+\delta)-\frac{k}{R^2}\label{eq:top-hatint}~,
\end{align}
where $a(t)$ and $R(t)$ are respectively the outer and inner scale factors and $\cdot$ denotes the time derivative. The density of the inner region is written in terms of the background density, $\bar{\rho}$, and the density contrast, $\delta$. It is customary to assume that the background space-time is spatially flat. Note that the spatial curvature of the inner space-time, $k$, cannot, in general be assumed to be a constant~\cite{Weinberg2003}. The energy densities inside and outside of the over-dense region each follow a conservation equation~:
\begin{align}
\frac{d}{dt}\bar{\rho}+3\frac{\dot{a}}{a}(\bar{\rho}(1+w)) &= 0~,\\
\frac{d}{dt}(\bar{\rho}(1+\delta))+3\frac{\dot{R}}{R}\left[\bar{\rho}(1+\delta)(1+w)\right] &= \Gamma\label{eq:dtrhoinGamma}~.
\end{align}
where the species is assumed to follow an equation of state of the form $p=w\rho$. The function $\Gamma$ is an adjustable parameter depending on wether the considered species is allowed to cluster. Many researches have argued that the quintessence component of the Universe should remain homogeneous at all time~\cite{Mota2004,Caldwell1997,Hwang2001}. This argument is based on the equation for the evolution of scalar field perturbations in the matter dominated era~\cite{Hwang2001},
\begin{equation}
\delta\ddot{\phi}+3H\delta\dot{\phi}+\left(k^2/a^2+V_{,\phi\phi}\right)\delta\phi=\dot{\phi}\dot{\delta}_m~,
\label{eq:dphi}
\end{equation}
where $\phi$, $\delta\phi$, $\delta_m$, $V$ and $V_{,\phi\phi}$ are respectively the quintessence scalar field, the field perturbation, the matter over-density, the scalar field potential and its second derivative w.r. to $\phi$. Eq.~\ref{eq:dphi} shows that there is a characteristic comoving mode $k_J/a\sim V_{,\phi\phi}$ below which the perturbations grow exponentially, corresponding to a typical length  $\lambda_J\sim 1/\sqrt{V_{\phi\phi}}$, sometimes referred to as the \emph{Jeans length}. As for most quintessential potentials, this length is larger than the Horizon and it is generally altogether assumed that quintessence does not cluster. However, the process preventing the clustering is not explained.

Returning to the top-hat equations, in the case where the energy density is that of a scalar field with potential $V$, Eq.~(\ref{eq:dtrhoinGamma}) reduces to 
\begin{equation}
\ddot{\phi}_{\tt loc}+3\frac{\dot{R}}{R}\dot{\phi}_{\tt loc}+\frac{dV}{d\phi}=\Gamma/\dot{\phi}_{\tt loc}~,
\label{eq:top-hat_Gamma}
\end{equation}
where $\phi_{\tt loc}$ is the local value of the field within the over-dense region. To prevent the field from collapsing, the phenomenological functional parameter $\Gamma$ must be set to~\cite{Mota2004}
\begin{equation}
\Gamma = -3\left(\frac{\dot{a}}{a}-\frac{\dot{R}}{R}\right)\dot{\phi}_{\tt loc}^2~.
\label{eq:Gammaquint}
\end{equation}
This makes the equation for the field inside the over-dense region strictly equivalent to that for the outside region thus forcing $\phi=\phi_{\tt loc}$. In practice, this amounts to demand that the field within the local region couples to the value taken by the expansion factor outside the region rather than the local $\frac{\dot{R}}{R}$. 

The top-hat model is purely phenomenological and not a rigorous solution of the equations of General Relativity. In presence of dust matter only, the spherically symmetric cosmological solution of Einstein equations is the \emph{Lema\^itre-Tolman-Bondi} (LTB) metric of which the top-hat model is a limit case~\cite{Rekier2015_1}. This model is not valid in the case of a general fluid with pressure momentum transfer. Such as is the case for quintessence.

The fully consistent relativistic treatment calls for numerical methods. Lasky et al. have made a step toward the generalisation of the LTB solution by formulating the problem as an initial value problem \cite{Lasky2006_2}. We follow another approach based on the many successes of Numerical Relativity.

\section{Formalism}
\label{sec:Formalism}
Following the work presented in~\cite{Rekier2015_1}, we write the spherically symmetric squared line-element as
\begin{equation}
ds^2=-(\alpha^2-\beta^2)dt^2+2\beta dtdr + \psi^4 a^2(t)(\hat{a}dr^2+\hat{b}r^2d\Omega^2)~,
\label{eq:ansatzBSSN}
\end{equation}
where $\alpha(t,r)$ is the lapse, $\beta(t,r)$ is the radial component of the shift $\beta_\mu$, and $\hat{a}$ and $\hat{b}$ denote the non-zero components of the diagonal conformal 3-metric. $\psi^2\sqrt{a}$ is the conformal factor. We have factored out the cosmological scale factor $a(t)$ which follows its own dynamics that serves as dynamical background. The extrinsic curvature is split into its trace $K$ and its conformally-scaled trace-free part $\hat{A}_{\mu\nu}$,
\begin{equation}
K_{ij} = \frac{1}{3}\gamma_{ij}K+\psi^4a^2\hat{A}_{ij}~,
\end{equation}
with $\gamma_{ij}$ being the spatial 3-metric. Due to spherical symmetry, $\hat{A}_{ij}$ has only two non-zero components $A_a:=\hat{A}^r_r$ and $A_b:=\hat{A}^\theta_\theta$. As $\hat{A}_{ij}$ is traceless, one further has $A_a+2A_b=0$. 
%The auxiliary 3-vector of the BSSN formalism only has one non-zero component \cite{Alcubierre2010is}, 
%\begin{equation}
%{\Delta}^r = \frac{1}{\hat{a}}\left[\frac{\partial_r\hat{a}}{2\hat{a}}-\frac{\partial_r\hat{b}}{\hat{b}}-\frac{2}{r}\left(1-\frac{\hat{a}}{\hat{b}}\right)\right]~.
%\end{equation}
The BSSN formalism ensures that $\det(\gamma_{ij})/(\phi^4a^3)^2=\det{(f_{ij})}$ at all time, where $f_{ij}$ is the flat metric which here translates to $\hat{a}\hat{b}^2=1$. 

For our purpose, we shall limit ourselves to the zero-shift gauge $\beta=0$. There is no formal difficulty in choosing a different gauge. This one however allows us to perform comparisons between cosmological models more straightforwardly. The normal vector which is tangent to the word-line of the Eulerian observer reads $n_\mu=(-\alpha,0,0,0)$. The set of dynamical and constraint equations was given in~\cite{Rekier2015_1}. %We repeat them here for convenience
%\begin{align}
%\partial_{t} \a &= -2\alpha \a A_{a},\nonumber\\
%\partial_{t} \b &=-2\alpha \b A_{b}, \nonumber\\
%\partial_{t} \psi &=-\frac{1}{6}\alpha \psi K - \frac{1}{2}\frac{\dot{a}}{a}\psi, \nonumber\\
%\partial_{t} K  & = - \nabla^{2}\alpha +
%\alpha(A_{a}^{2} + 2A_{b}^{2} + \frac{1}{3}K^{2}) \nonumber \\
%    & + 4\pi\alpha(E+S_{a}+2S_{b}),\nonumber\\
%\partial_{t} A_{a} & = - \left(\nabla^{r}\nabla_{r}\alpha - \frac{1}{3}\nabla^{2}\alpha\right)
%+ \alpha\left(R^{r}_{r} - \frac{1}{3}R\right) \nonumber \\
%   & + \alpha K A_{a} - \frac{16\pi}{3}\alpha(S_a - S_b)\nonumber\\
%\partial_{t}\hat {\Delta}^{r} & = - \frac{2}{\a}(A_{a}\partial_{r}\alpha + \alpha\partial_{r}A_{a}) + 2\alpha\left(A_{a}\hat{\Delta}^{r} - \frac{2}{r\b}(A_{a}-A_{b})\right)\nonumber \\
%    &+ \frac{\xi \alpha}{\a} \left[\partial_{r}A_{a} - \frac{2}{3}\partial_{r}K + 6A_{a}\frac{\partial_{r}\psi}{\psi} \right.\nonumber\\
%    & \left.+(A_{a}-A_{b})\left(\frac{2}{r}+\frac{\partial_{r}\b}{\b}\right) - 8\pi j_{r} \right],
%\end{align}
%We fix $\xi=2$ as this is the standard BSSN optimal choice and guarantees strong hyperbolicity. 
The source terms are 
\begin{align}
	E&:=n_{\alpha}n_{\beta}~T^{\alpha\beta}~, \nonumber \\
	j_{i}&:=-\gamma_{i\alpha}~n_{\beta}~T^{\alpha\beta}~, \nonumber \\
	S_{ij}&:=\gamma_{\alpha i}\gamma_{\beta j}~T^{\alpha\beta}~,
\label{eq:sourceterms}
\end{align}
where $E$ is the energy density as seen by the Eulerian observer, $j_i$ is the momentum transfer and $S_{ij}$ is the stress tensor. In spherical symmetry and using adapted coordinates~:
\begin{align}
S_{ij} =
 \begin{pmatrix}
S_{rr} & & \\
& S_{\theta\theta} & \\
& & S_{\theta\theta}
 \end{pmatrix}~,
 &&
j_i = (j_r(t,r),0,0)~.
\end{align}
We further define $Sa:=S^r_r$ and $Sb:=S_\theta^\theta$.
The dynamics of the quintessence field is given by the Klein-Gordon equation 
\begin{equation}
\partial_\mu\left((-g)^{1/2}\partial^\mu\phi\right)=(-g)^{1/2}\partial_\phi V~,
\end{equation}
where  $g=\det{(g_{\mu\nu})}$, the determinant of the complete metric. In order to write this as a first-order system, one defines \cite{Torres2014}
\begin{align}
\Pi &:= n_\mu\partial^\mu\phi~,\label{eq:defPi}\\
\Psi_i &:= \partial_i\phi~.\label{eq:defPsi}
\end{align}
Assuming spherical symmetry, the evolution equations for the quintessence field and the related variables \ref{eq:defPi} and \ref{eq:defPsi} can be written as (using the short-hand $\Psi\equiv\Psi_r$)
\begin{align}
\pounds_n\phi &=\Pi~,\\
\pounds_n\Psi &=\frac{1}{\alpha}\partial_r(\alpha\Pi)~,\\
\pounds_n\Pi &= K\Pi+\frac{1}{\alpha}D_r(\alpha\Psi)-\frac{dV}{d\phi}~.
\end{align}
In presence of both matter and scalar field, the energy source functions have two components. The expressions for the scalar field components read
\begin{align}
E_\phi &=\frac{1}{2}\left(\Pi^2+\frac{\Psi^2}{\psi^4a^2\a}\right)+V~,\label{eq:E_phi}\\
S^\phi_a &=\frac{1}{2}\left(\Pi^2+\frac{\Psi^2}{\psi^4a^2\a}\right)-V~,\\
S^\phi_b &=\frac{1}{2}\left(\Pi^2-\frac{\Psi^2}{\psi^4a^2\a}\right)-V~,\\
j_r^\phi &= -\Pi\Psi\label{eq:jr_phi}~.
\end{align}
In the zero shift gauge, the evolution equations reduce to (making use of $\a\b^2=1$).
\begin{align}
\partial_t\Pi &= \alpha K \Pi - \frac{1}{\psi^6a^2r^2}\partial_r\left(\alpha\frac{\psi^2r^2}{\a}\Psi\right)-\alpha\frac{dV}{d\phi}\label{eq:dtPi}~,\\
\partial_t\Psi &= \partial_r(\alpha\Pi)~,\\
\partial_t\phi &= \alpha\Pi~.
\end{align}
%The Hamiltonian and momentum constraints read
%\begin{align}
%  \mathcal{H} & \equiv R - (A^{2}_{a} + 2A^{2}_{b})+ \frac{2}{3}K^{2}-16\pi E = 0~, \\
%  \mathcal{M}^r &\equiv \partial_{r}A_{a} - \frac{2}{3}\partial_{r}K + 6A_{a}\frac{\partial_{r}\psi}{\psi}\nonumber\\
%  & +(A_a -A_b)\left(\frac{2}{r} + \frac{\partial_r \b}{\b}\right) - 8\pi j_r = 0~.
%\end{align}
The numerical solution for the dynamics is obtained in a way similar to what was presented in~\cite{Rekier2015_1} using a second-order \emph{Partially-Implicit-Runge-Kutta} (PIRK) method~\cite{Cordero-Carrion2012,Cordero2014}; more details can be found in the appendix to this paper. The radial dimension is approximated by a uniformly discretised cell-centred grid, and radial derivatives are computed with a fourth-order finite difference scheme. We use fourth-order Kreiss-Oliger dissipation. A few virtual points of negative radius are added to the numerical grid to ensure that the numerical profiles have the correct parity throughout the integration.

The entire code used to produce the simulations presented in this work has been made publicly available on the web at \texttt{http://github.com/jrekier/FORTCosmoSS}.

The evolution of the background follows the Friedmann and acceleration equations
\begin{align}
\frac{1}{\alpha_{\tt bkg}^2}\left(\frac{\adot}{a}\right)^2 &= \frac{8\pi}{3} \bar{\rho},
\label{eq:Friedmann}\\
\frac{1}{\alpha_{\tt bkg}^2}\frac{\addot}{a}-\frac{\dot{a}}{a}\frac{\dot{\alpha}_{\tt bkg}}{\alpha_{\tt bkg}}&= -\frac{8\pi}{6}\bar{\rho}(1+3w),
\label{eq:acc}
\end{align}
where $\alpha_{\tt bkg}$ is the background value of the lapse function.

We impose radiative conditions at the outer boundary
\begin{equation}
\partial_t f = \partial_t f_{\tt bkg} - v \, \partial_r f - \frac{v}{r} (f-f_{\tt bkg}),
\end{equation}
where $v$ is the speed of propagation of the variable $f$ on the grid. This is inferred by considering the characteristic structure of the variables of the evolution system of equations. $f_{\tt bkg}(t)$ denotes the spatially homogeneous asymptotic cosmological value of the variable $f$ and $\partial_t f_{\tt bkg}$ its first time derivative.

We work in natural units with $G=c=1$. In order for the complete set of units to be dimensionless, we further impose the value of the Hubble factor measured today to be equal to some adjustable parameter, $H_0=x t_\text{\tt scale}^{-1}$. Comparison with the experimental value $\sim 70$km/s/Mpc fixes the time scale. The length and mass scales are then obtained from $l_\text{\tt scale}=c t_\text{\tt scale}$ and $m_\text{\tt scale}=(c^3/G)t_\text{\tt scale}$. One can dispose of the need to specify the particular set of scales employed within a computation by expressing these in terms of $H_0$.

We consider two different dark energy models besides the simplistic $\Lambda$CDM. The first one is the inverse power-law \emph{Ratra-Peebles} (RP) model~\cite{Ratra:1987rm} characterised by the potential 
\begin{equation}
V(\phi)=\frac{M^{4+n}}{\phi^n}~,
\end{equation}
where $n$ and $M$ are constants. This produces late time cosmological acceleration. In the general case where the field is not at rest initially, the slow-roll conditions are not satisfied at initial time as the field starts with a small value corresponding to a steep region of the potential. When it is present, the dust matter density dominates over the energy density and the Universe assumes a power-law expansion in time, $a\sim t^{2/3}$, during which $\phi$ rolls down its potential. The field eventually comes to the flat tail region of the potential where the weak energy condition ($w<-1/3$) is violated leading to a de Sitter expansion phase. If the field is initially at rest, the evolution proceeds in the same way but starts with another de Sitter phase. 

The other quintessence model considered is the \emph{Pseudo-Nambu-Goldstone Boson} model (PNGB)~\cite{Frieman:1995pm}. While the RP model is an example of \emph{freezing} model where the violation of the weak energy condition happens at late time, the PNGB is classified within the set of \emph{thawing} models in which this violation happens at early times \cite{amendola2010}. The analytical expression for the PNGB potential is 
\begin{equation}
V(\phi)=\mu^4\cos\left(1+\phi/f\right)~.
\end{equation}

\section{Scalar field evolution}
\label{sec:FormalismSF}

In order to test the stability and convergence properties of our method, we start by considering the evolution of a spatial distribution of scalar field in a way similar to what was performed for the analysis of the gauge dynamics in Rekier et al.~\cite{Rekier2015_1}. The purpose of the background dynamics of this section is merely to allow us to test our method on a de Sitter background and is by no means to reproduce the measured expansion. The initial quintessence profile is~:
\begin{align}
&\phi = \phi_{\tt bkg}(1+\delta\phi)~,\\
&\delta\phi(t=0,r) = \frac{\delta\phi_0~ r^2}{1+r^2}\left[e^{-\frac{(r-r_i)^2}{\sigma_i^2}}+e^{\frac{(r-r_i)^2}{\sigma_i^2}}\right]~,
\end{align}
where $\delta\phi_0$ sets the initial amplitude of the pulse, $r_i$ its initial position and $\sigma_i$ its spatial extension. The background value is chosen as the solution of
\begin{equation}
H_i^2=\frac{8\pi}{3}V(\phi_{\tt bkg})n,
\end{equation}
which is just the Friedmann equation with $\alpha=1$. The initial data are set after imposing~:
\begin{align}
\a(t=0)&=\b(t=0)=1~,\\
K(t=0) &= -3H_i~;~A_a(t=0) = A_b(t=0) = 0~.
\end{align}
This reduces the Hamiltonian constraint to 
\begin{equation}
a^{-2}\psi^{-5}(\partial_r^2\psi+\frac{2}{r}\partial_r\psi)+6H_i^2=16\pi E_\phi~,
\label{eq:intitialBVPphi}
\end{equation}
which is solved at initial time as a boundary value problem. 
We perform two simulations each one corresponding to a different value of the initial expansion factor. These are performed with the RP potential with $n=2$. The mass scale $M$ is chosen in order for the field to reproduce the behaviour of a cosmological constant when the fields has a value around $\phi_0\sim\sqrt{8\pi}$ \cite{amendola2010}:
\begin{equation}
8\pi V(\phi_0)=\Lambda~.
\end{equation}
The initial amplitude of the scalar field inhomogeneity parameters are $\delta\phi_0=5 \times 10^{-4}$, $\sigma=2$ and $r_i=20$. The field is assumed to be initially at rest ($\Pi(t=0)=0$). Eq.~(\ref{eq:intitialBVPphi}) then turns into 
\begin{equation}
a^{-2}\psi^{-5}(\partial_r^2\psi+\frac{2}{r}\partial_r\psi)+6H_i^2 = 8\pi\left(\frac{\Psi^2}{\psi^4a^2\a}\right)+16\pi V(\phi)~.
\end{equation}

\begin{figure}
  \begin{center}
    \includegraphics[width=0.5 \textwidth]{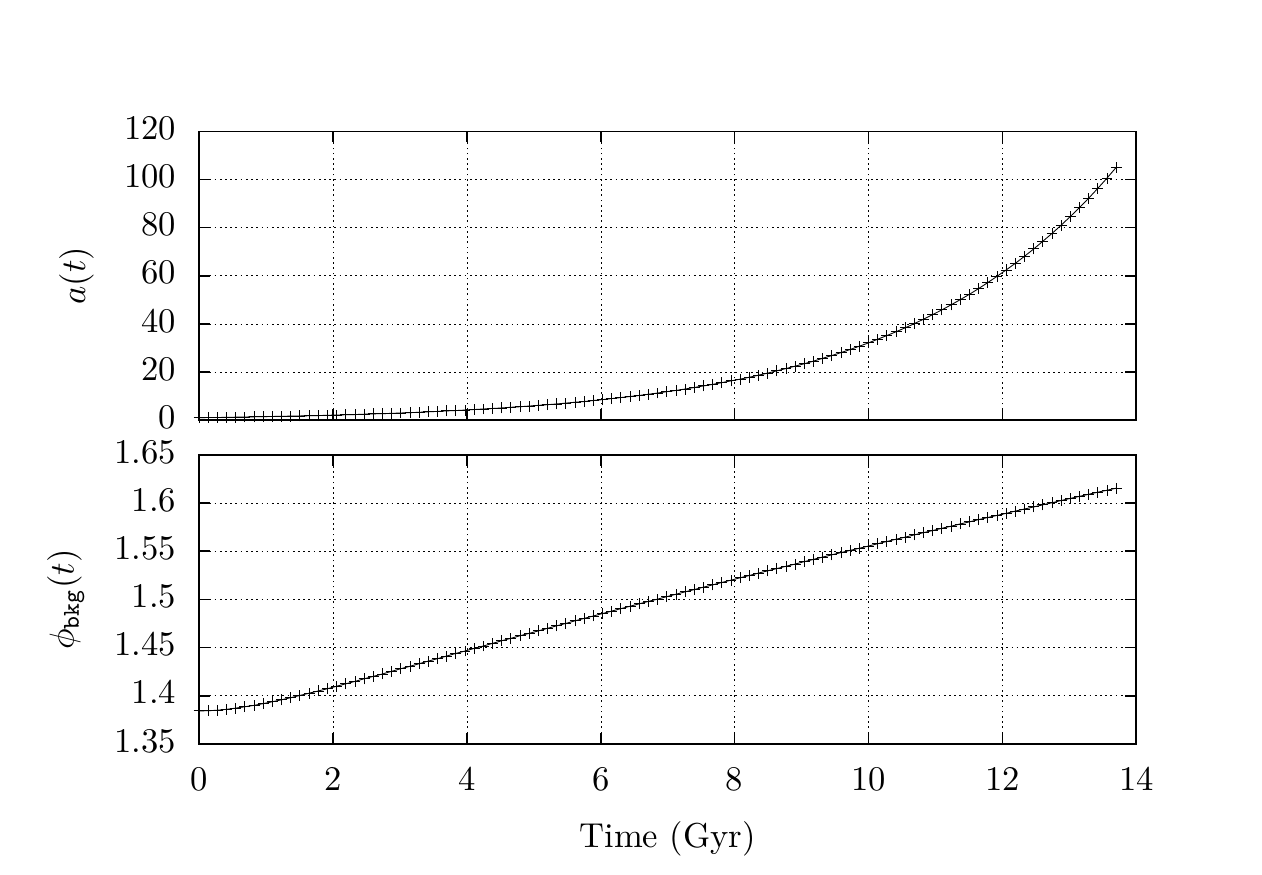}
  \end{center}
  \caption{Long-term cosmological evolution of the scale factor and background scalar field for a Universe filled with quintessence ($H_i=5H_0$).}
  \label{fig:cosmo_scalar_pulse1}
\end{figure}

\begin{figure}
  \begin{center}
    \includegraphics[width=0.5 \textwidth]{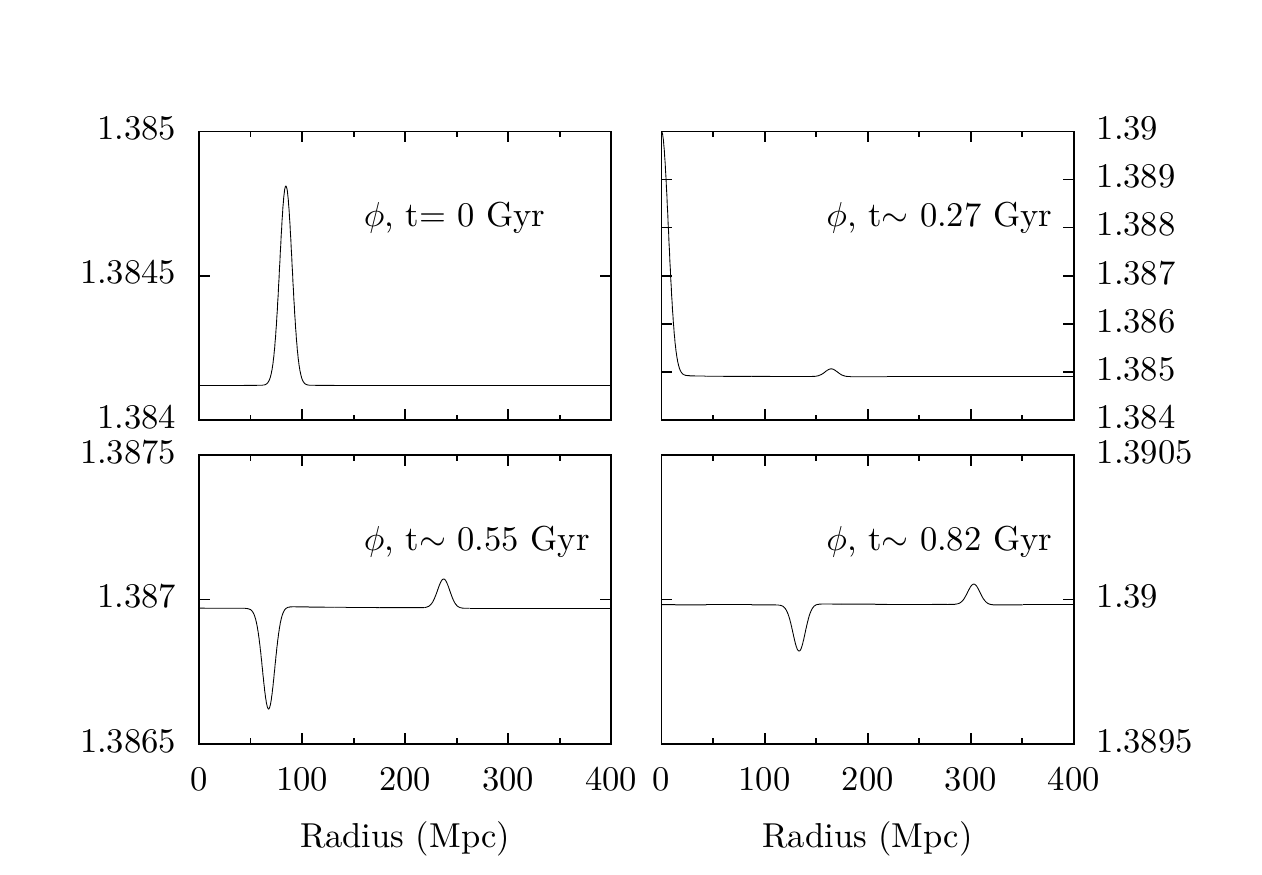}
  \end{center}
  \caption{Evolution of a gaussian quintessence scalar pulse profile on a de Sitter background ($H_i=5H_0$).}
  \label{fig:phi_pulse_geod1}
\end{figure}

We first study the evolution of the field in the case where the initial expansion factor is of the order of the present day Hubble factor ($H_i=5H_0$). The evolution of the scale factor and the homogeneous part of the scalar field are shown on Fig.~\ref{fig:cosmo_scalar_pulse1}.
The Universe starts off in a phase of de Sitter expansion as the slow-roll conditions are met at initial time. The field rapidly unfreezes as it rolls down its potential leading to a milder expansion rate before eventually freezing out again in the tail of its potential. The scalar pulse propagation happens within the early de Sitter phase. The potential part of the field energy density dominates over the term proportional to the gradient of the scalar field $\Psi$ at initial time. The evolution of the scalar field profile in the geodesic slicing is shown on Fig.~\ref{fig:phi_pulse_geod1}.
The scalar pulse separates into two parts. The inward travelling pulse gets reflected from the origin of coordinates and then travels outward. The apparent dynamics is very similar to what we encountered in Ref.~\cite{Rekier2015_1} in the study of a gauge pulse. However, the physical situation here is very different as we are now dealing with a non-vanishing distribution of energy. The situation here can be seen as the evolution of a spherical shell initially placed at a radius $r_i$. After the shell has bounced from the origin, the central value of the field returns to its homogeneous asymptotical value leaving no effect on the local expansion. 

\begin{figure}
  \begin{center}
    \includegraphics[width=0.5 \textwidth]{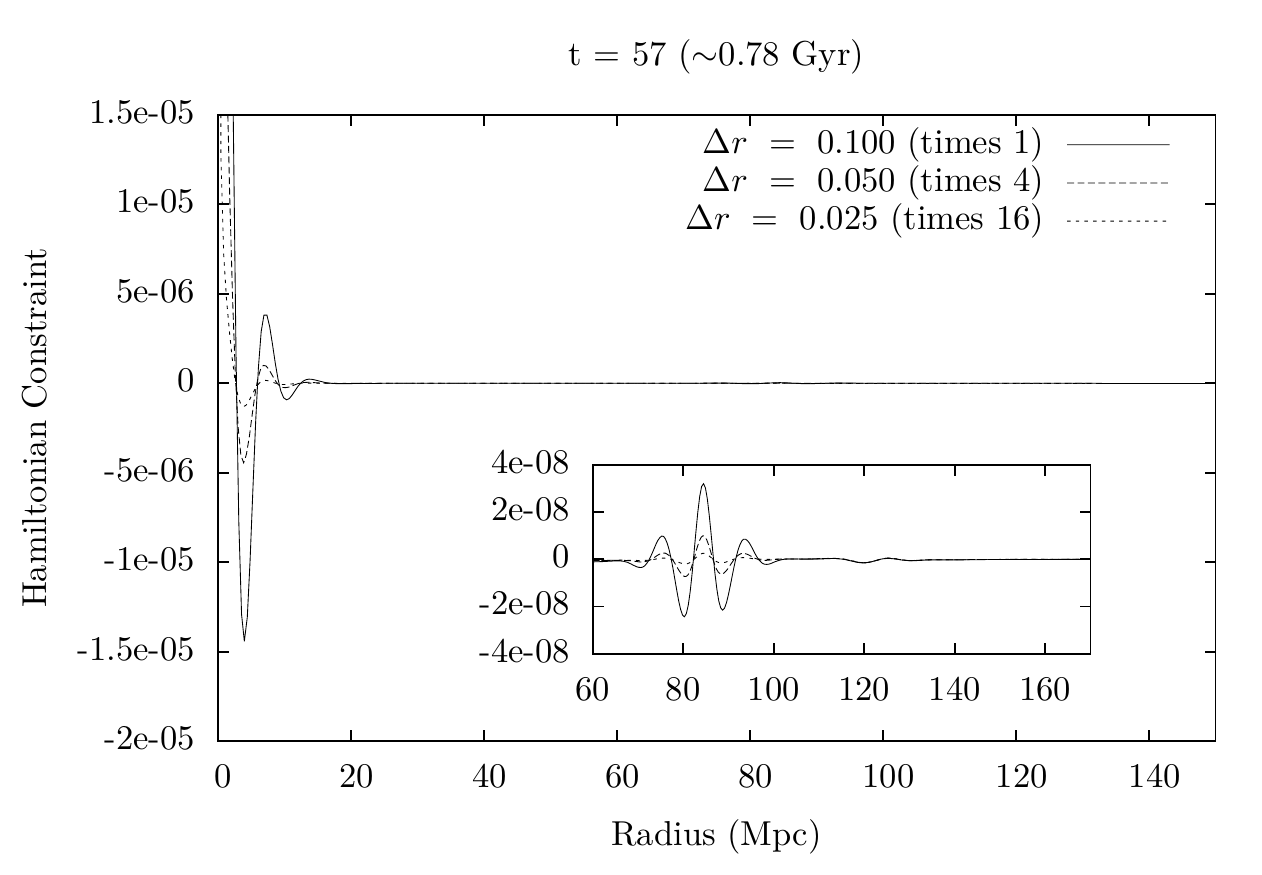}
  \end{center}
  \caption{Hamiltonian constraint violation profile resulting from the propagation of a gaussian pulse in geodesic slicing ($H_i=5H_0$).}
  \label{fig:H_scalar_pulse_geod_1}
\end{figure}

\begin{figure}
  \begin{center}
    \includegraphics[width=0.5 \textwidth]{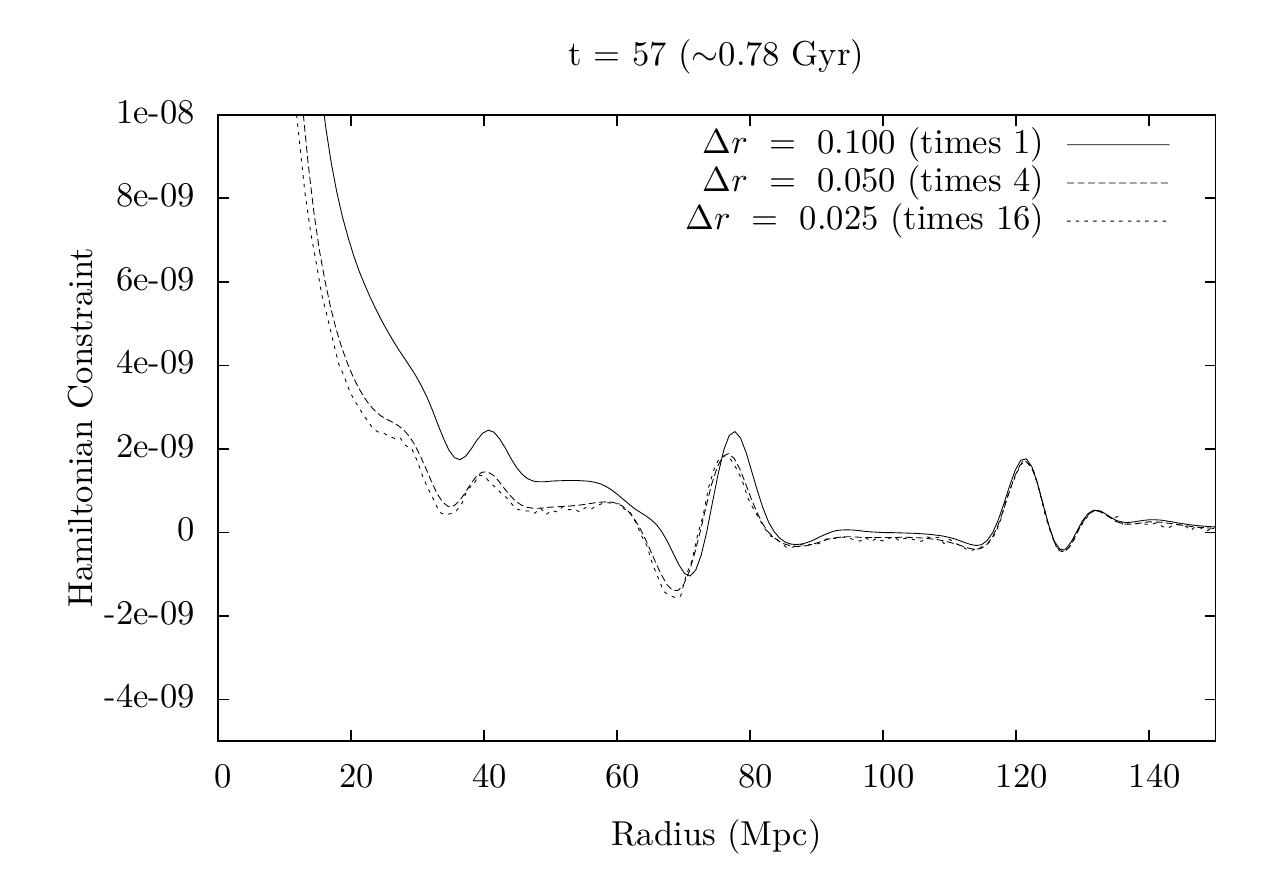}
  \end{center}
  \caption{Hamiltonian constraint violation profile resulting from the propagation of a gaussian pulse in Bona-Masso slicing ($H_i=5H_0$).}
  \label{fig:H_scalar_pulse_BonaMasso}
\end{figure}

The violation of the Hamiltonian constraint profile is shown on Fig.~\ref{fig:H_scalar_pulse_geod_1} for the evolution in the geodesic slicing gauge and for 3 values of the resolution. The inner plot shows a close-up of the outer plot around the radius corresponding to the initial position of the scalar field pulse. A similar plot is shown for the result of the evolution in the Bona-Masso slicing with $f=0.333$ on Fig.~\ref{fig:H_scalar_pulse_BonaMasso}. This choice of gauge is made in order for the coordinate speed of light to remain finite throughout the integration~\cite{Torres2014}. Incidentally, setting $f=1/3$ makes the Bona-Masso slicing equivalent to the conformal time gauge. In both slicings, the convergence of the method is beyond second-order. The error is maximal at the centre of coordinates but remains controlled throughout the integration.

\begin{figure}
  \begin{center}
    \includegraphics[width=0.5 \textwidth]{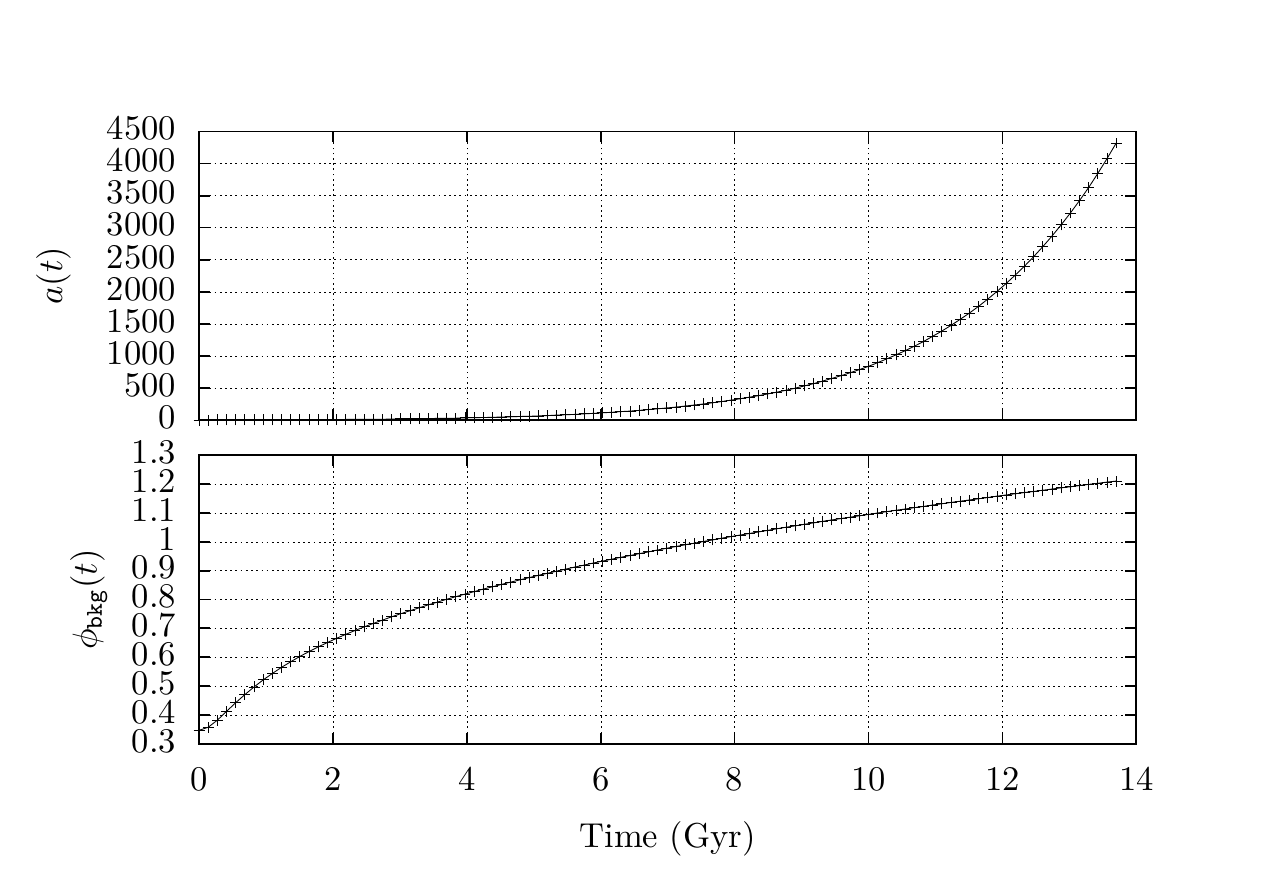}
  \end{center}
  \caption{Long-term cosmological evolution of the scale factor and background scalar field for a Universe filled with quintessence ($H_i=20H_0$).}
  \label{fig:cosmo_scalar_pulse2}
\end{figure}

\begin{figure}
  \begin{center}
    \includegraphics[width=0.5 \textwidth]{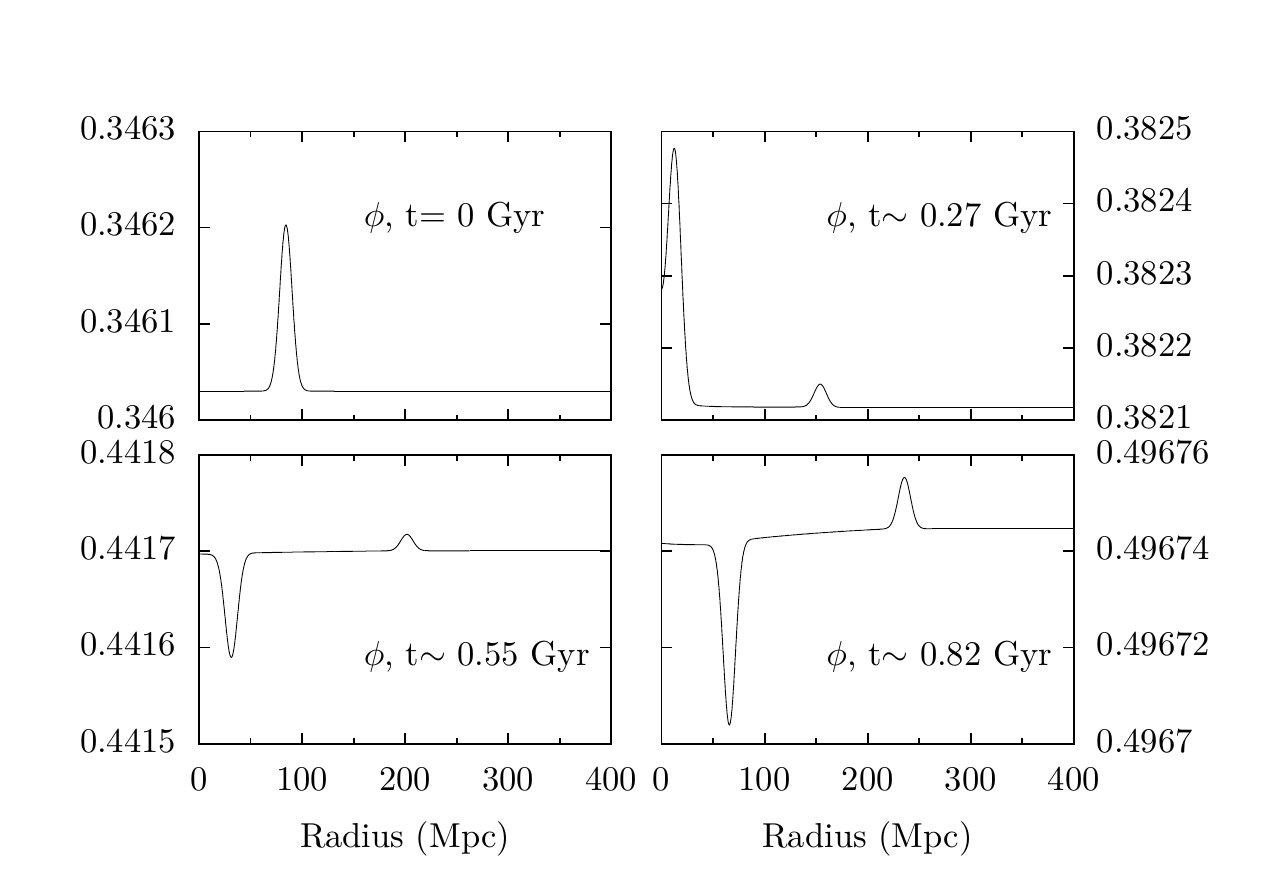}
  \end{center}
  \caption{Evolution of a gaussian quintessence scalar pulse profile on a de Sitter background ($H_i=20H_0$).}
  \label{fig:phi_scalar_pulse_geod2}
\end{figure}

\begin{figure}
  \begin{center}
    \includegraphics[width=0.5 \textwidth]{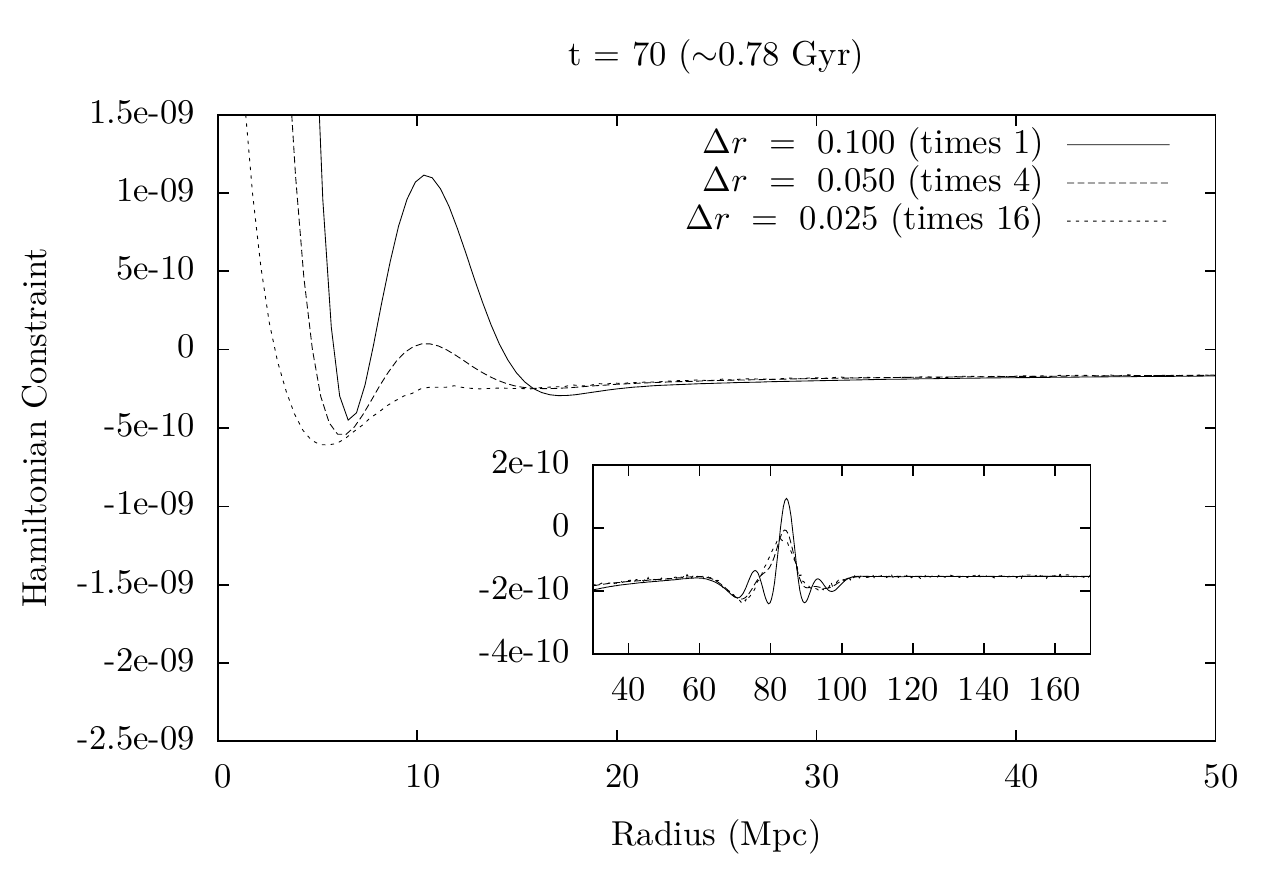}
  \end{center}
  \caption{Hamiltonian constraint violation profile resulting from the propagation of a gaussian pulse in geodesic slicing ($H_i=20H_0$).}
  \label{fig:H_scalar_pulse_geod_2}
\end{figure}

We now study the case where the initial expansion factor is $H_i=20H_0$, that is one order of magnitude larger than the Hubble factor today. The evolution of the scale factor and the homogeneous asymptotical value of the scalar field are shown on Fig.~\ref{fig:cosmo_scalar_pulse2}. The milder expansion between the two de Sitter phases happens earlier. As opposed to the previous case, the dominant part of the initial energy density is proportional to the gradient $\Psi$ which in fact leads to a value of the density smaller than its asymptotical value. Figure~\ref{fig:phi_scalar_pulse_geod2} shows the evolution of the pulse on the spatial domain. The central value of the field does not return to its asymptotical value after the pulse is reflected. We have used geodesic slicing.
The Hamiltonian violation profile is shown on Fig.~\ref{fig:H_scalar_pulse_geod_2}. The convergence is again above second-order.

\begin{figure}
  \begin{center}
    \includegraphics[width=0.5 \textwidth]{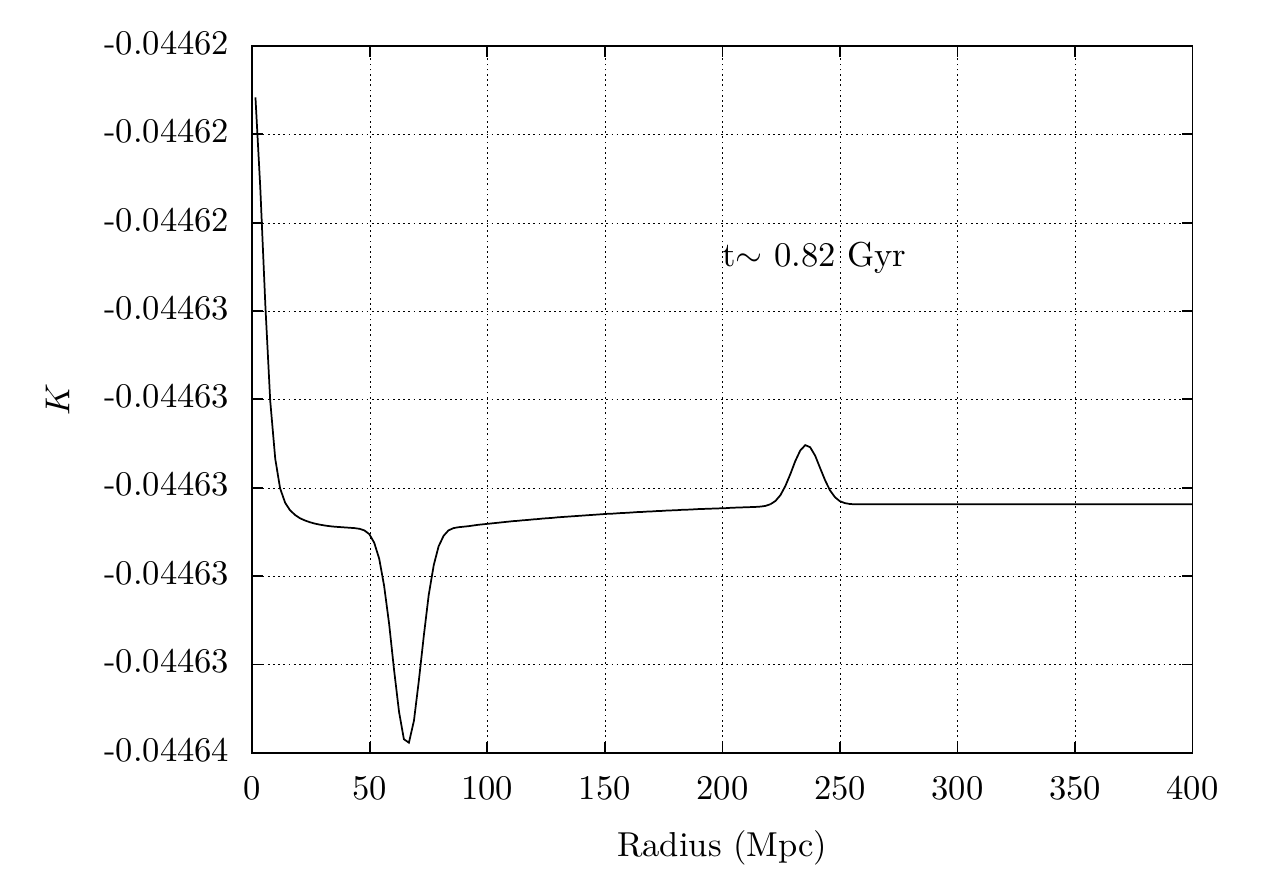}
  \end{center}
  \caption{Trace of the extrinsic curvature resulting from the propagation of a scalar pulse on a de Sitter background ($H_i=20H_0$).}
  \label{fig:K_scalar_pulse_geod_2}
\end{figure}

The fact that the central value of the field is different from the asymptotical value at late-time has an impact on the local expansion around the centre of coordinates. Fig.~\ref{fig:K_scalar_pulse_geod_2} shows the trace of the extrinsic curvature tensor which is proportional to the opposite of the local isotropic expansion of space-time after the pulse has been reflected from the centre. This shows how the expansion is smaller than that of the background at the centre of coordinates causing the local expansion to fall behind that of the background thus creating a spherical space-time inhomogeneity region of mildest expansion around the centre of coordinates, an effect that is persistent and remains visible after the pulse is reflected. 

\section{Comparison with the top-hat}
\label{sec:compTH}
\begin{figure*}
  \begin{center}
    \includegraphics[width=0.7 \textwidth]{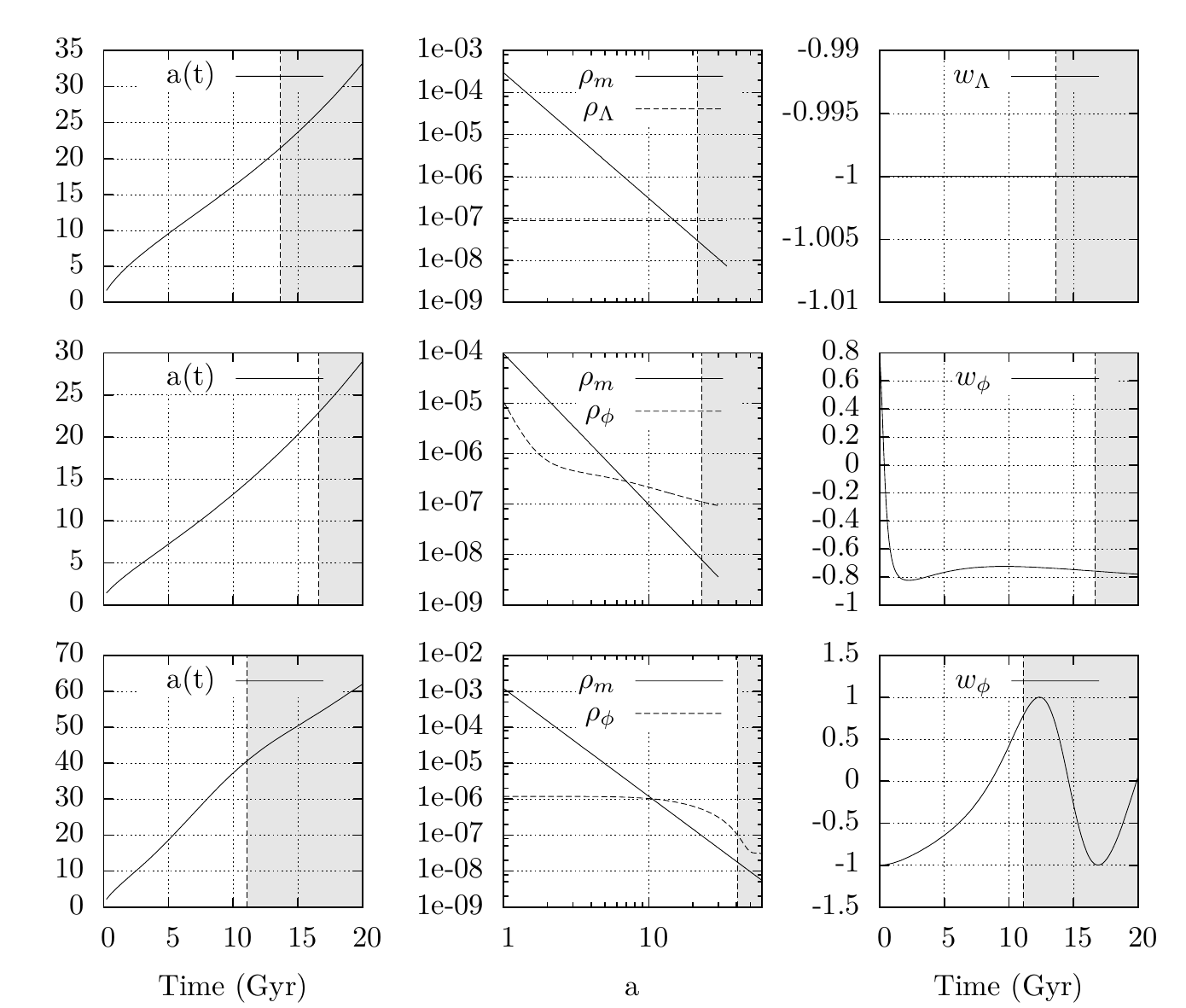}
  \end{center}
  \caption{Evolution of the three background models considered for the study of spherical collapse in presence of quintessence.}
  \label{fig:3-models_bkg}
\end{figure*}

We now turn to the comparison of our method to the top-hat model of spherical collapse. This is done for $\Lambda$CDM as well as the 2 quintessence models discussed earlier. The background evolution are chosen as qualitatively similar to the evolution of our Universe yet, as the fitting of the models onto the observed data is not our primary concern, the quintessence models parameters are chosen in order to display a significant departure from the $\Lambda$CDM model. Fig.~\ref{fig:3-models_bkg} shows the evolutions of the scale factor, the components of the energy density and the equation of state parameter of the 3 considered backgrounds.
The initial value of the scale factor is $a=1$ in all models. The vertical dashed line corresponds to the point of zero redshift where the value of the Hubble factor equals the value measured today. The shaded regions of each plot corresponds to times of negative redshift. The quantities $\rho_m$, $\rho_\Lambda$ and $\rho_\phi$ correspond to the value of the background energy density respectively associated to dust, cosmological constant and quintessence. These are given in units in which $H_0=10^{-3}$. $w_m$, $w_\Lambda$ and $w_\phi$ denote the corresponding equation of state parameters.

The cosmological parameters used for each model are shown in Table~\ref{tab:cosm_param_bkg}.
\begin{table}\centering
\ra{1.3}
\begin{tabular}{l*3l}    
\toprule
\emph{Model} & $\Lambda$CDM & RP & PNGB   \\
\midrule
$\Omega_m^i$ & 0.9997  & 0.9 & 0.999   \\ 
$\Omega_\Lambda^i$ & $1-\Omega_m^i$ & 0 & 0 \\ 
$\Omega_\phi^i$ & 0 & $1-\Omega_m^i$ & $1-\Omega_m^i$ \\
$H_i$ & $50H_0$ & $30H_0$ & $100H_0$ \\
$w_{\Lambda/\phi}^i$ & $-1$ & $0.8$ & $-1$ \\\bottomrule
 \hline
\end{tabular}
\caption{\label{tab:cosm_param_bkg} Initial cosmological parameters employed to produce the simulations of Fig.~\ref{fig:3-models_bkg}.}
\end{table}
The model-specific parameters for the RP potential are
\begin{align}
n &= 2~,\nonumber\\
M &= \frac{2}{5}\left(\frac{3H_0^2}{8\pi}\right)^{1/(4+n)}(8\pi)^{\frac{n}{8+n}}~.
\end{align}
The parameters of the PNGB model are
\begin{align}
f &= 15\pi/\sqrt{8\pi}~,\nonumber\\
\mu &= 6\left(\frac{3H_0^2}{8\pi}\right)^{1/4}~.
\end{align}
These are chosen in order to produce acceleration at times close to the present day.

The comparison between the fully relativistic solution obtained by solving the complete set of BSSN equations and the top-hat solution is first studied by looking at the central value of the local scale factor defined as $a\psi^2(t,r=0)$ in the BSSN coordinates. The quintessence field is assumed to be homogeneous at initial time. The only perturbation to the homogeneous background comes from the matter energy density. We write the initial matter density contrast, $\delta_m:=\left(\rho_m/\bar{\rho}_m-1\right)$ with $\bar{\rho}_m$ being the background matter density, as a step-like function in order to be close to the distribution of matter in the top-hat model.
\begin{equation}
\delta_m(t=0,r)=\delta_m^0\left(\frac{1}{2}-\frac{1}{2}\tanh(k\left(r-r_{\tt span})\right)\right)~,
\end{equation}
where $\delta_m^0$ is the value of the density contrast at the centre of coordinates and $r_{\tt span}$ is the radius at which it drops to half of its maximum value. The parameter $k$ adjusts the steepness of the profile. 

We take $r_{\tt span}\sim 20$Mpc, that is of the order of the size of superclusters formed at late time in the History of the Universe. This can be considered as small compared to the size of the Horizon yet sufficiently large to consider the inner value of the energy density to be homogeneous. As we have pointed out, the quintessence component should cluster very little on such scales. Our method allows to test this \emph{ab initio}. We compare the results of our computation with two realisations of the top-hat model in which the clustering parameter, $\Gamma$, is zero corresponding to a case where the quintessence is fully allowed to cluster (later referred to as ``Top-Hat w.c.'') and when it takes the form of Eq.~(\ref{eq:Gammaquint}) allowing no clustering at all (later referred to as ``Top-Hat n.c.''). We use geodesic slicing with zero shift thus allowing a direct comparison of the metric components and ensuring identical time coordinates in both models.

\begin{figure}
  \begin{center}
    \includegraphics[width=0.49 \textwidth]{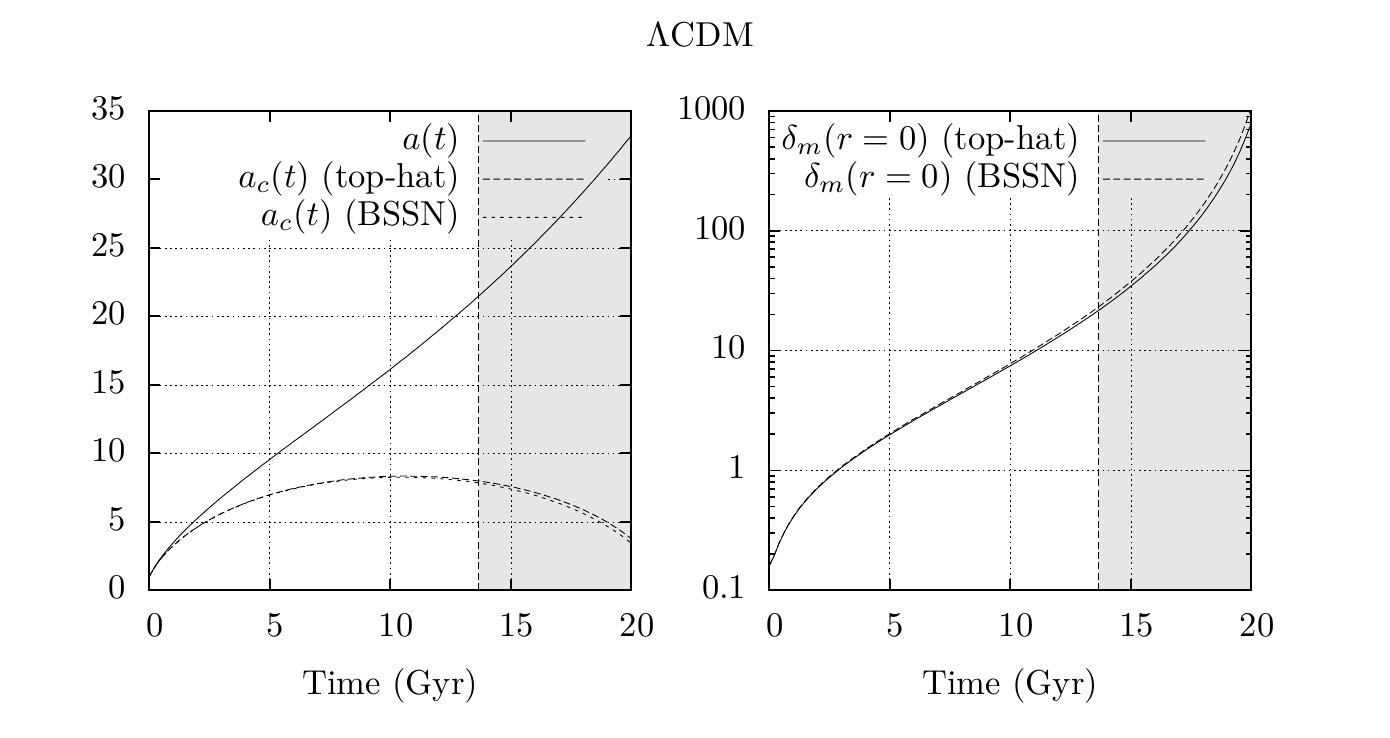}
  \end{center}
  \caption{Evolution of the scale factor and the density contrast for the fully relativistic solution and the top-hat solutions in the $\Lambda$CDM model.}
  \label{fig:collapse_LCDM}
\end{figure}

\begin{figure}
  \begin{center}
    \includegraphics[width=0.49 \textwidth]{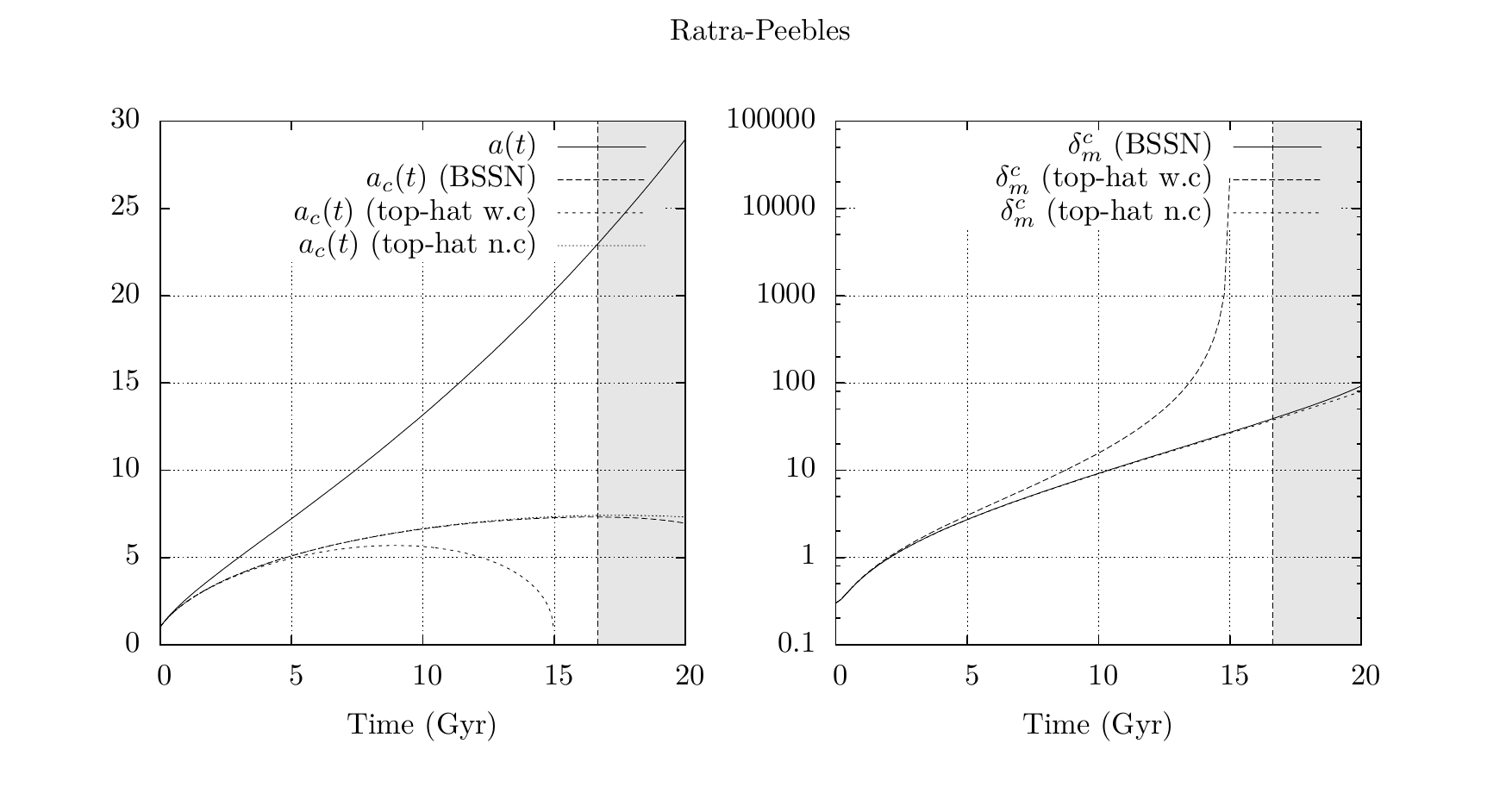}
  \end{center}
  \caption{Evolution of the scale factor and the density contrast for the fully relativistic solution and the top-hat solution in the Ratra-Peebles model.}
  \label{fig:collapse_RP}
\end{figure}

\begin{figure}
  \begin{center}
    \includegraphics[width=0.49 \textwidth]{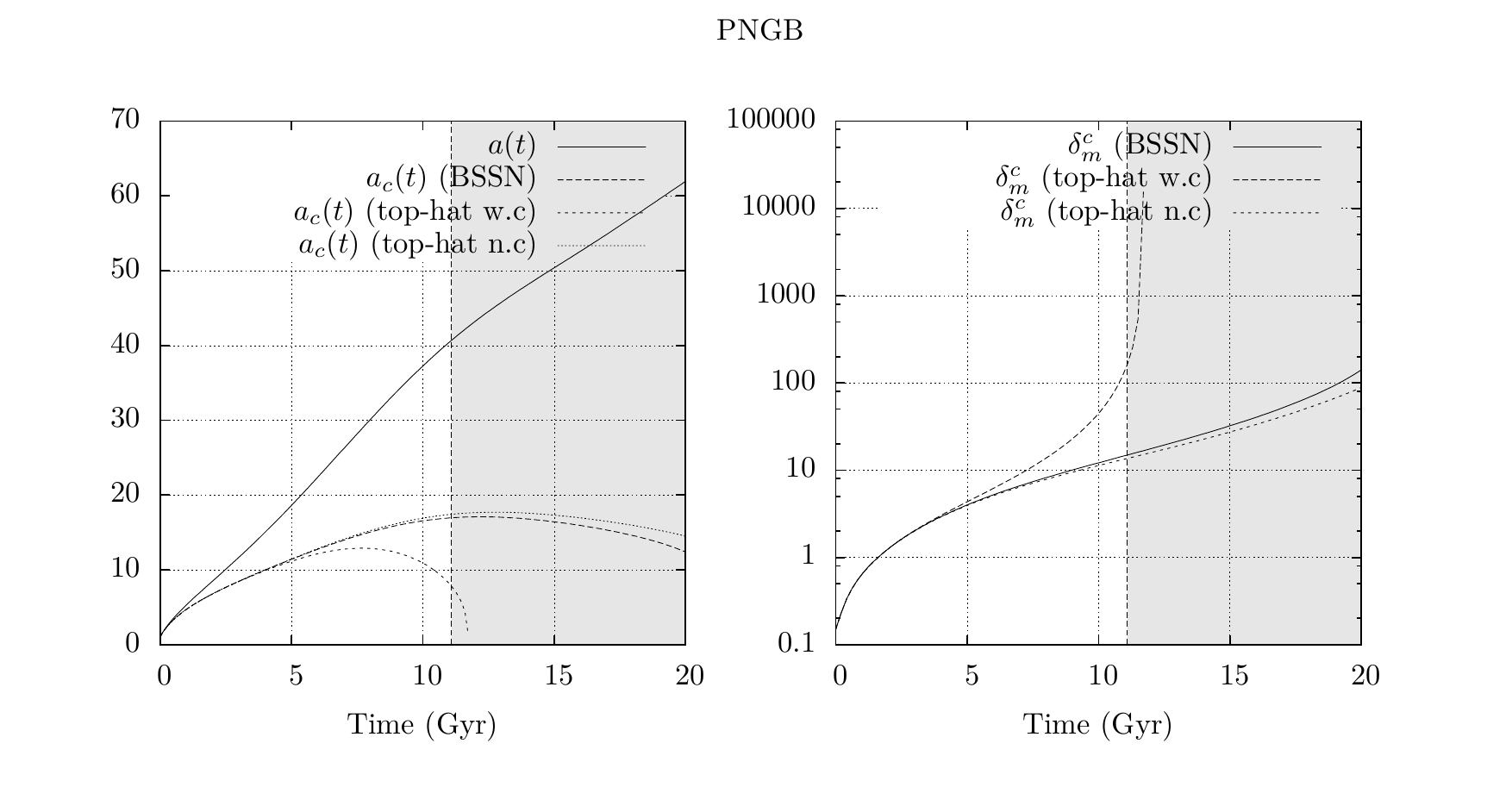}
  \end{center}
  \caption{Evolution of the scale factor and the density contrast for the fully relativistic solution and the top-hat solution in the PNGB model.}
  \label{fig:collapse_PNGB}
\end{figure}

The evolution of the background and central values of the scale factor for the $\Lambda$CDM model are shown on Fig.~\ref{fig:collapse_LCDM} along with the density contrast at the centre of coordinates. The initial value of the latter being fixed to $\delta_m(r=0)=0.16$.
In spite of the conceptual limitations of the top-hat model, this reproduces the correct relativistic predictions remarkably as far as the scale factor and contrast density are concerned. The correspondance is close to being exact up to very deep within the non-linear growth regime of $\delta_m$. The same quantities are shown on Fig.~\ref{fig:collapse_RP} and Fig.~\ref{fig:collapse_PNGB} for the Ratra-Peebles and PNGB models respectively with an initial density contrast $\delta_m=0.3$ and $\delta_m=0.15$.
The evolution of the scale factor and density contrast is really close to both top-hat solution at early time when the matter dominates over the quintessence field. When quintessence starts to dominate, the naive top-hat model with clustering predicts a quicker collapse than the relativistic solution does. This solution remains close to the top-hat solution without clustering at all time. The small departure from the top-hat model that is observed at late time for the PNGB potential could be an effect of a small yet non-zero late-time clustering of quintessence. 

\begin{figure}
  \begin{center}
    \includegraphics[width=0.5 \textwidth]{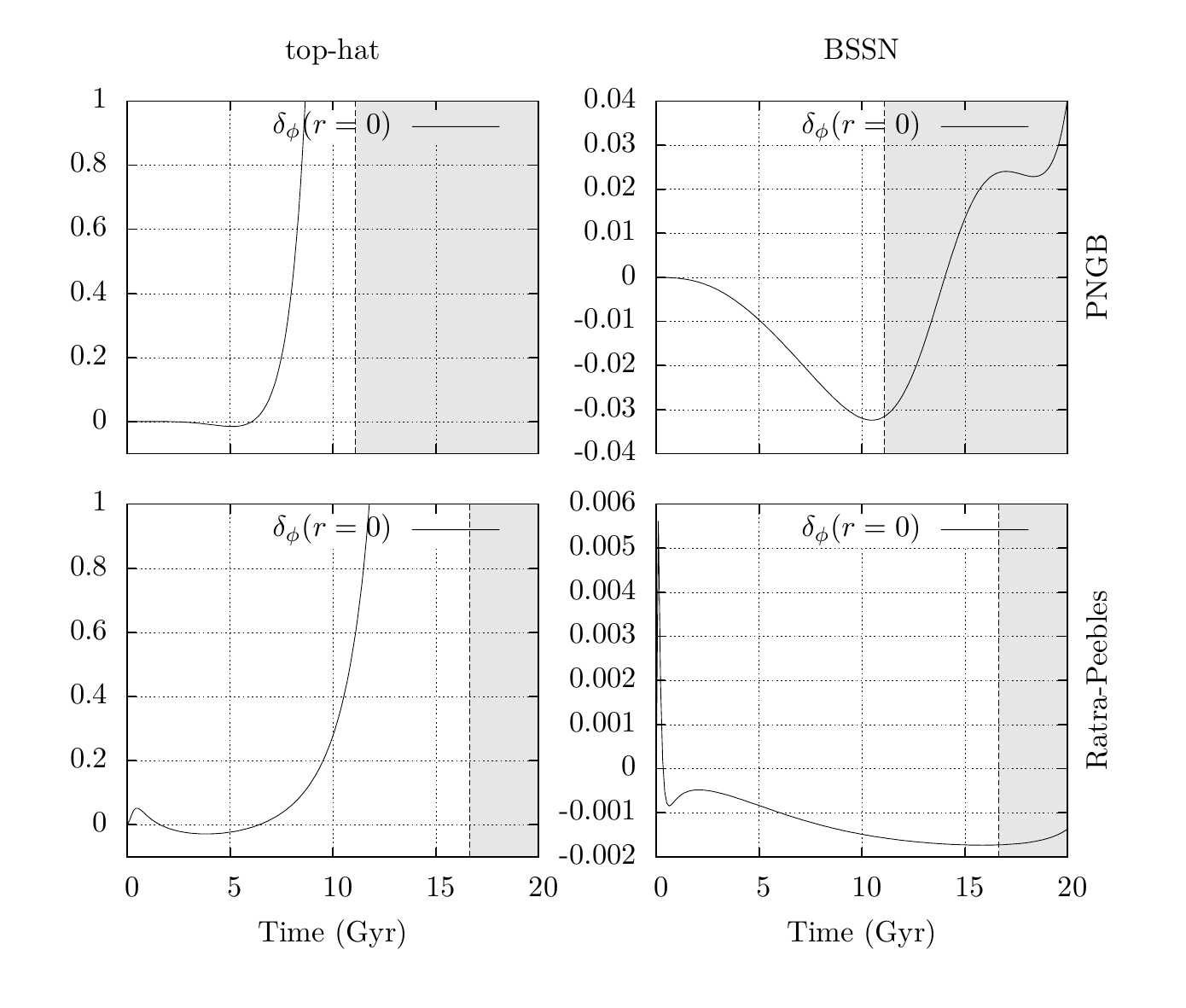}
  \end{center}
  \caption{Evolution of the central value of the quintessence energy density during the collapse for 2 models of quintessence.}
  \label{fig:2-models-delta_phi}
\end{figure}

Fig.~\ref{fig:2-models-delta_phi} shows the central value of the quintessence density contrast, $\delta_\phi:=\left(\rho_\phi/\bar{\rho}_\phi-1\right)$, with $\bar{\rho}_\phi$, for the top-hat model with clustering of quintessence and for the BSSN simulation for both models of quintessence. The top row corresponds to the PNGB model, the second row correspond to a Ratra-Peebles model. No \emph{ad hoc} assumption is made regarding the Jeans length of the scalar field. We can see that the growth of the quintessence over-densities reach its non-linear regime while remaining bounded yet non-zero.

\section{Beyond the top-hat model}
\label{sec:beyond}

The method that we have developed allows to explore the physical reasons behind the fact that quintessence clusters very little.
The top-hat model, regardless of wether it allows quintessence to cluster, assumes that the over-dense region has the symmetries of the FLRW space-time. This forbids, in particular, the existence of anisotropic pressures and momentum transfers. While this is not a problem when only dust matter is present, this is rather arbitrary in presence of quintessence. In the latter case, anisotropic pressure terms do build up values that are comparable to the inhomogeneous part of the isotropic pressure. The latter are themselves kept very small however for other reasons. 

In this section, we investigate the evolution of anisotropies during the collapse process for the PNGB and Ratra-Peebles models.

Spherical symmetry forbids any anisotropic quantities to be along directions other than the radial one. The momentum transfer associated to a scalar field is given by Eq.~(\ref{eq:jr_phi}). It is useful to think of the scalar field as a non-perfect fluid in order to identify the radial anisotropic pressure term as 
\begin{equation}
\pi_\phi:=\pi^\phi_{rr}=(S_a^\phi-S_b^\phi)~.
\end{equation}
This comes out as $\pi_\phi=\frac{\Psi^2}{\psi^4a^2\hat{a}}$. It is zero in the background space-time, as it should be due to the vanishing of quintessence gradient.

\begin{figure}
  \begin{center}
    \includegraphics[width=0.5 \textwidth]{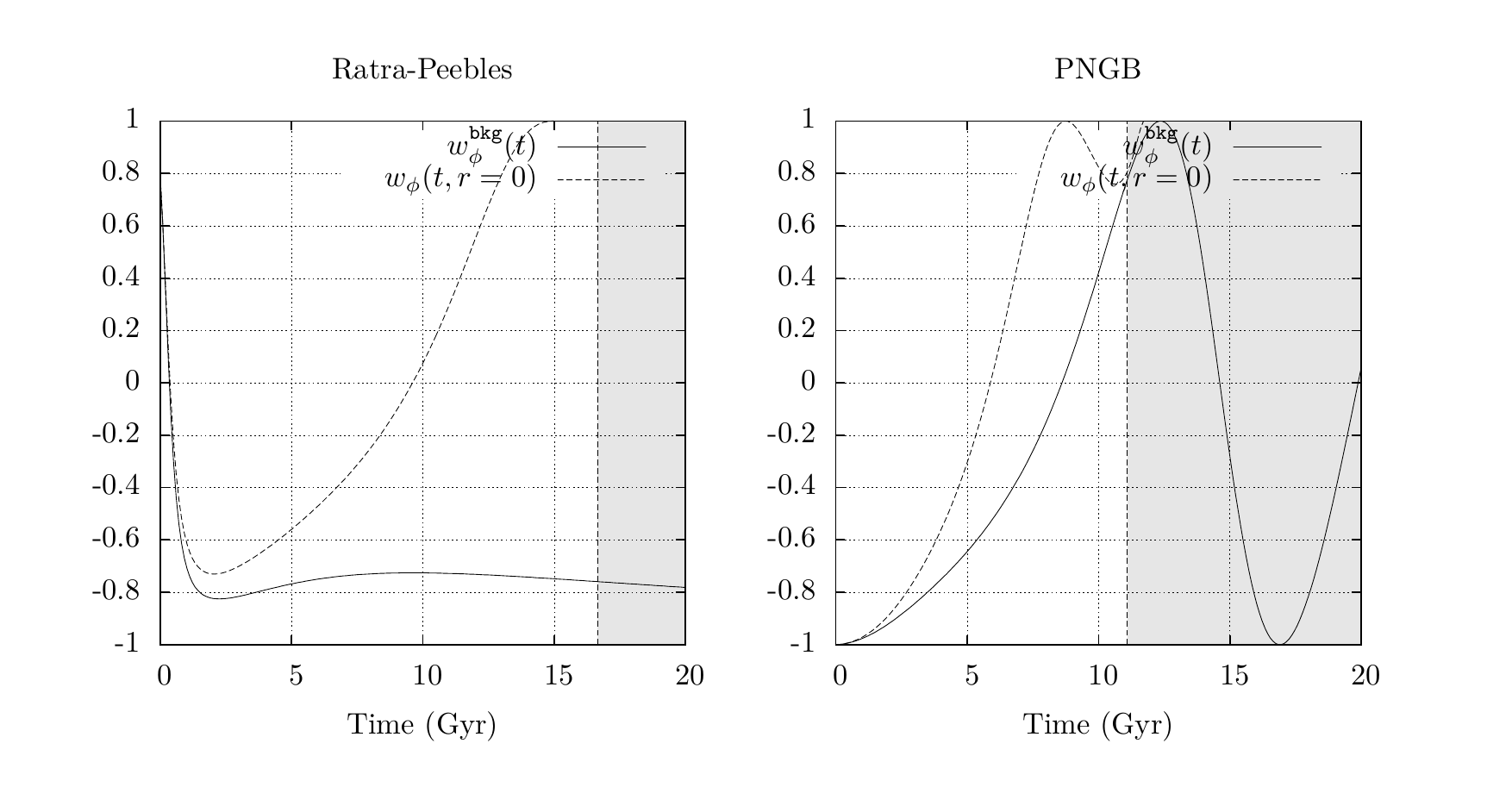}
  \end{center}
  \caption{Evolution of the central and background values of the equation of state parameter in the Ratra-Peebles and PNGB models in the naive top-hat picture in which quintessence fully clusters.}
  \label{fig:naive_tp_w}
\end{figure}

In order to understand why the field does not cluster, it is best to consider why it should collapse in the naive top-hat model ($\Gamma=0$) in the first place. As it turns out, most of the difference in the energy density of the field comes from a difference in kinetic energy. Fig.~\ref{fig:naive_tp_w} shows the evolution of the effective quintessence equation of state parameter inside and outside of the over-dense region in this model for both the RP and PNGB models. 
This raises very quickly at early time but this has a limited impact as dust matter is still the dominant component of the energy density. The corresponding increase in the quintessence kinetic energy is reinforced at the turnover when the over-dense region stops expanding and starts to collapse. This can be understood by considering Eq.~(\ref{eq:dtPi}) describing the evolution of the field momentum which, in the  spatially homogeneous case reduces to 
\begin{equation}
\partial_t\Pi = K\Pi - \frac{dV}{d\phi}~.
\end{equation}
A collapsing space corresponds to $K>0$ leading to a positive feedback on the growth of $\Pi$ which can hardly be counterbalanced by the gradient of the potential, especially when the latter is very flat such as is the case in slow-roll expansion. The amount of kinetic energy built up within the over-dense region causes the quintessence field to act as stiff matter which accelerates the collapse. 

The faulty part of this picture lies in the fact that the two parts of space-time are completely disjoined and there is no possibility of momentum transfer between both regions. This transfer is made possible in the complete picture through the second term of Eq.~(\ref{eq:dtPi}). The modified top-hat model with no clustering reproduces this coupling artificially through the term proportional to $\Gamma$. This induces a loss of momentum proportional to the difference between the extrinsic curvatures of the inner and outer regions of space that effectively compensates the positive feedback effect described above. In reality, this term is purely phenomenological but is also non-local and Eq.~(\ref{eq:top-hat_Gamma}) does not correspond to a solution of the equations of General Relativity.

The method that we have developed to solve for the complete relativistic dynamics allows to investigate what actually happens at the local scale. 

\begin{figure*}
  \begin{center}
    \includegraphics[width=1 \textwidth]{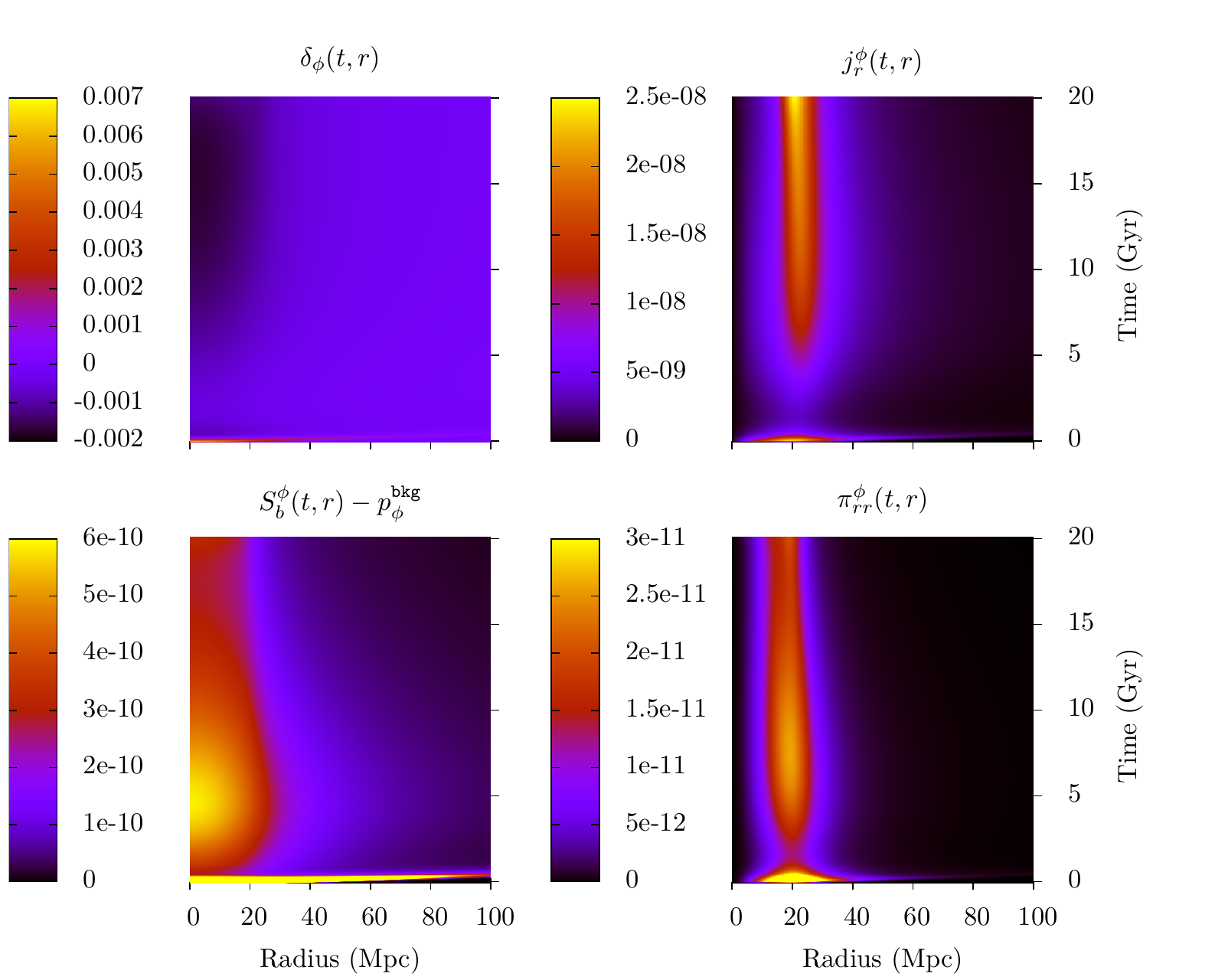}
  \end{center}
  \caption{Evolution of the anisotropies in the Ratra-Peebles model}
  \label{fig:anisotropies_RP}
\end{figure*}

\begin{figure*}
  \begin{center}
    \includegraphics[width=1 \textwidth]{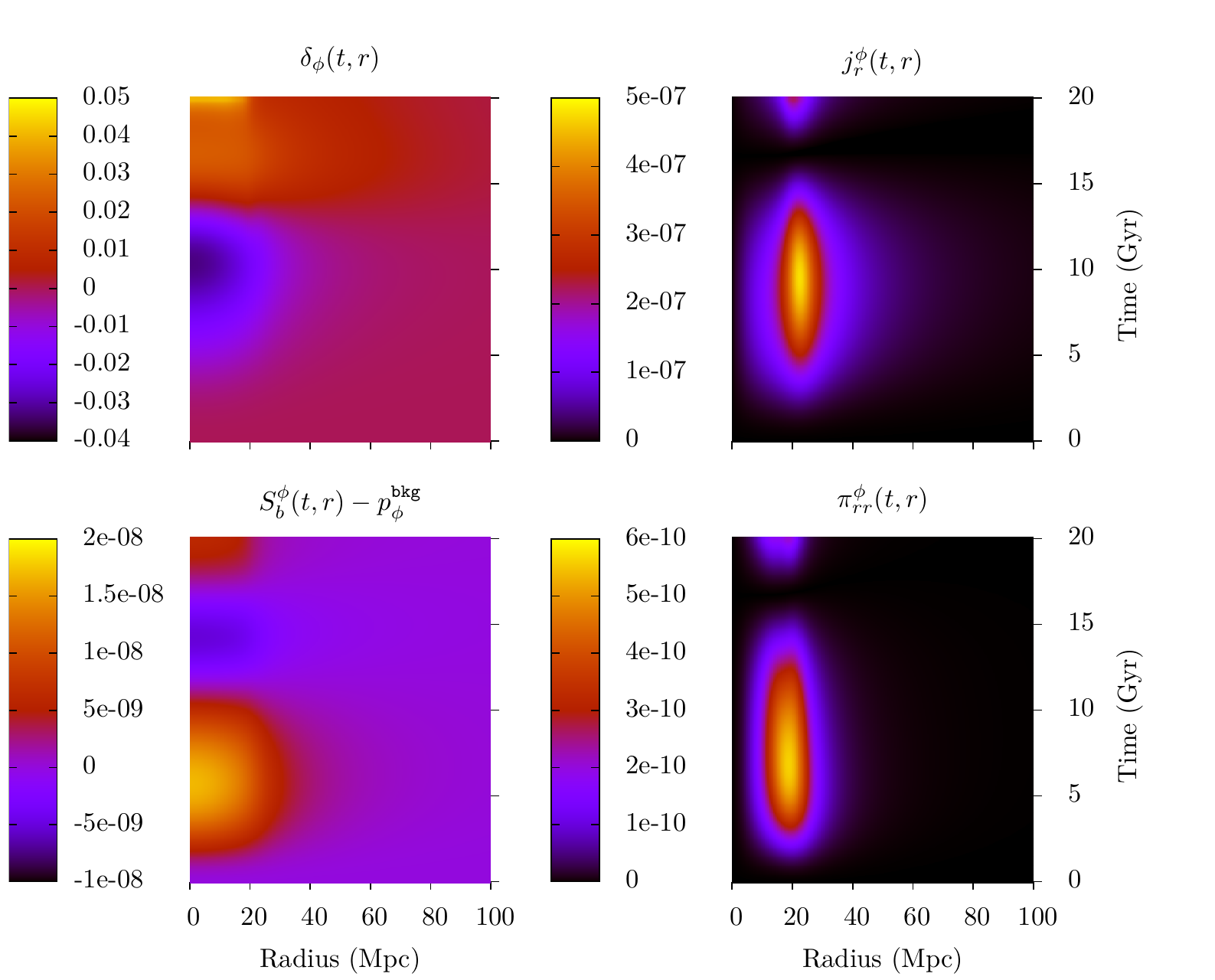}
  \end{center}
  \caption{Evolution of the anisotropies in the PNGB model}
  \label{fig:anisotropies_PNGB}
\end{figure*}

The upper rows of Fig.~\ref{fig:anisotropies_RP} and Fig.~\ref{fig:anisotropies_PNGB} show the evolution of the field density contrast profile as a function of time along with the radial momentum transfer from the stress-energy tensor $j_r^\phi$ respectively for the Ratra-Peebles and the PNGB models. These figures were obtained with the same parameters as in the previous section. The results are presented in natural units with the Hubble constant value set to $H_0=0.001$.

As the over-dense region gets closer to the turn-over, the momentum transfer raises to a large positive value corresponding to an outward transfer. This is responsible for balancing the kinetic energy inside and outside of the over-dense region. This transfer is maximal around the boundary of the over-dense region which is the place were the gradients of scalar field and metric variables are maximal.

The decrease in the momentum of the field adjusts the equation of state parameter at the centre of coordinates to its background value. As the Universe draws more into the vacuum dominated era, this may either result in a more rapid or delayed collapse depending on the shape of the potential. The PNGB model has a positive equation of state parameter at late-time resulting in a facilitated collapse which in turns increases the small value of the quintessence density contrast. In both models, this small yet non-zero density contrast induces a small difference of pressure inside the over-dense region. The bottom rows of Fig.~\ref{fig:anisotropies_RP} and Fig.~\ref{fig:anisotropies_PNGB} show the evolutions of the gradient of isotropic pressure between the over-dense and background regions (left panel) and the anisotropic pressure profiles (right panel). One sees that the isotropic pressure value dominates over the isotropic pressure yet the latter is one order of magnitude less than the anisotropic pressure around the boundary of the over-dense region where it is maximal.

For completeness, we ought to provide an explanation for the colours in the plots of Fig.~\ref{fig:anisotropies_RP} at early time. These are a consequence of the transient behaviour following the evolution of the particular set of chosen initial conditions. At initial time, the field starts off in a very steep region of its potential. This results in a large gain of momentum at the centre of coordinates which is not yet counterbalanced by the negative feedback of expansion nor the momentum transfer that is initially null. The small increase in $\delta_\phi$ that results, along with a positive equation of state $w_\phi \sim 1$ lead to a temporary high pressure difference that overshoots the colour scale of the plot. The effect is rapidly counterbalanced by expansion and momentum transfer. The small initial ``jump'' in the contrast density is perceivable on the two bottom plots of Fig.~\ref{fig:2-models-delta_phi}. One understands how the contrast density gets down slower in the top-hat picture with clustering as the only effect is the negative feedback from initial expansion. These transients behaviours have little effect on the overall evolution as these happen at a time when the matter energy density is dominant. These however encourages us to look for more general initial conditions.

%\section{lipsum}
%\lipsum

\section{Conclusion and Perspectives}
\label{section:conclusion}

In this work we have proved the validity of a new method for solving the equations of evolution of a spherically symmetric cosmological space-time filled with a real quintessence scalar field. We have proved the stability and validity of our method by studying the evolution of a spherical distribution of quintessence inhomogeneity on a de Sitter background. This allowed us to study the deviation of the local expansion around quintessence over-densities in formation and opens perspectives of more detailed study of the backreactions of local distributions of energy on the cosmological expansion.

We have shown how the method can be used to study the impact of quintessence on the spherical collapse of a matter over-density. We have undertaken this study for 3 different cosmological backgrounds and compared our results to those obtained in the top-hat picture. It turns out that the top-hat model predicts evolutions of the scale factor relatively close to the solution obtained using the fully relativistic method when the quintessence field is artificially kept homogeneous. The solution diverge slightly at the end of the integration when small over-densities do build up. We have identified the cause of the overall great homogeneity of the quintessence field as the existence of a non-null momentum transfer and anisotropic pressure component of the stress-energy tensor that are maximum at the boundary of the spherical over-density. 

We see the present work as a first step towards a more systematic study of the impact of the shape of the quintessence potential on the physics of galaxy clusters. It would be interesting to look at the evolution of geodesics near the boundary of the forming object. Even though the anisotropic terms are small, we expect that these might have an effect in the context of scalar-tensor theories where these are expected to couple directly to space-time curvature. In most treatments of spherical collapse, the top-hat model is assumed to be valid up to a time when it is assumed that the structure gets virialised. The top-hat model does not allow to account for the process and the argument of virilisation is used as a workaround and not as a prediction. We did not need this assumption in the present work. A generalisation of our method to varieties of energy with more general equations of state would be an interesting task that could ultimately allow to investigate the physics of the very late stages of the evolution, just before the collapse. 

\begin{acknowledgments}
J.~R. is supported by a FRS-FNRS (Belgian Fund for Scientific Research) Research Fellowship.  A.~F. is partially supported by the ARC convention No. 11/15-040. I. C.-C. acknowledges support from the European Research Councill (ERC) through the Starting Independent Research Grant CAMAP-259276. This work was also supported by the Spanish Government AYA2013-40979-P) and the local Autonomous Government (Generalitat Valenciana, grant Prometeo-II/2014/069). Computations were performed at the ``plate-forme technologique en calcul intensif'' (PTCI) of the University of Namur, Belgium, with the financial support of the FRS- FNRS (conventions No. 2.4617.07. and No. 2.5020.11). 
\end{acknowledgments}

%Use \appendix* if there only one appendix.
\appendix
\section{PIRK operator splitting for scalar field equations}
\label{section:appPIRK}
The operator splitting for the space-time quantities is the same as the one used in~\cite{Rekier2015_1}. The $\Psi$ function is evolved explicitly. The $\Pi$ function is evolved partially-implicitly using the following splitting:
\begin{align}
	&L_{2(\Pi)} = \frac{\alpha}{a^2\psi^4\a}\Psi\left(\frac{2}{r}-\frac{\partial_r\a}{\a}+\frac{\partial_r\alpha}{\alpha}+2\frac{\partial_r\psi}{\psi}\right), \\
	&L_{3(\Pi)} = \alpha K \Pi + \frac{\alpha}{a^2\psi^4\a}\partial_r\Psi-\alpha\frac{dV}{d\phi}~.
\end{align}
These quantities are evolved prior to the auxiliary BSSN quantity $\hat{\Delta}^r$, the evolution scheme of which remains unchanged.

\bibliographystyle{apsrev}
\bibliography{bibfile,bibfile_DE}

\begin{thebibliography}{39}
\expandafter\ifx\csname natexlab\endcsname\relax\def\natexlab#1{#1}\fi
\expandafter\ifx\csname bibnamefont\endcsname\relax
  \def\bibnamefont#1{#1}\fi
\expandafter\ifx\csname bibfnamefont\endcsname\relax
  \def\bibfnamefont#1{#1}\fi
\expandafter\ifx\csname citenamefont\endcsname\relax
  \def\citenamefont#1{#1}\fi
\expandafter\ifx\csname url\endcsname\relax
  \def\url#1{\texttt{#1}}\fi
\expandafter\ifx\csname urlprefix\endcsname\relax\def\urlprefix{URL }\fi
\providecommand{\bibinfo}[2]{#2}
\providecommand{\eprint}[2][]{\url{#2}}

\bibitem[{\citenamefont{Rekier et~al.}(2015)\citenamefont{Rekier,
  Cordero-Carri\'on, and F\"uzfa}}]{Rekier2015_1}
\bibinfo{author}{\bibfnamefont{J.}~\bibnamefont{Rekier}},
  \bibinfo{author}{\bibfnamefont{I.}~\bibnamefont{Cordero-Carri\'on}},
  \bibnamefont{and} \bibinfo{author}{\bibfnamefont{A.}~\bibnamefont{F\"uzfa}},
  \bibinfo{journal}{Phys. Rev. D} \textbf{\bibinfo{volume}{91}},
  \bibinfo{pages}{024025} (\bibinfo{year}{2015}),
  \urlprefix\url{http://link.aps.org/doi/10.1103/PhysRevD.91.024025}.

\bibitem[{\citenamefont{Riess et~al.}(1998)}]{Riess:1998cb}
\bibinfo{author}{\bibfnamefont{A.~G.} \bibnamefont{Riess}} \bibnamefont{et~al.}
  (\bibinfo{collaboration}{Supernova Search Team}), \bibinfo{journal}{Astron.
  J.} \textbf{\bibinfo{volume}{116}}, \bibinfo{pages}{1009}
  (\bibinfo{year}{1998}), \eprint{astro-ph/9805201}.

\bibitem[{\citenamefont{Perlmutter et~al.}(1999)}]{Perlmutter:1998np}
\bibinfo{author}{\bibfnamefont{S.}~\bibnamefont{Perlmutter}}
  \bibnamefont{et~al.} (\bibinfo{collaboration}{Supernova Cosmology Project}),
  \bibinfo{journal}{Astrophys. J.} \textbf{\bibinfo{volume}{517}},
  \bibinfo{pages}{565} (\bibinfo{year}{1999}), \eprint{astro-ph/9812133}.

\bibitem[{\citenamefont{Riess et~al.}(2001)}]{Riess:2001gk}
\bibinfo{author}{\bibfnamefont{A.~G.} \bibnamefont{Riess}} \bibnamefont{et~al.}
  (\bibinfo{collaboration}{Supernova Search Team}),
  \bibinfo{journal}{Astrophys. J.} \textbf{\bibinfo{volume}{560}},
  \bibinfo{pages}{49} (\bibinfo{year}{2001}), \eprint{astro-ph/0104455}.

\bibitem[{\citenamefont{Kowalski et~al.}(2008)}]{Kowalski:2008ez}
\bibinfo{author}{\bibfnamefont{M.}~\bibnamefont{Kowalski}} \bibnamefont{et~al.}
  (\bibinfo{collaboration}{Supernova Cosmology Project}),
  \bibinfo{journal}{Astrophys. J.} \textbf{\bibinfo{volume}{686}},
  \bibinfo{pages}{749} (\bibinfo{year}{2008}), \eprint{0804.4142}.

\bibitem[{\citenamefont{Amanullah et~al.}(2010)}]{Amanullah:2010vv}
\bibinfo{author}{\bibfnamefont{R.}~\bibnamefont{Amanullah}}
  \bibnamefont{et~al.}, \bibinfo{journal}{Astrophys. J.}
  \textbf{\bibinfo{volume}{716}}, \bibinfo{pages}{712} (\bibinfo{year}{2010}),
  \eprint{1004.1711}.

\bibitem[{\citenamefont{Komatsu et~al.}(2009)}]{Komatsu:2008hk}
\bibinfo{author}{\bibfnamefont{E.}~\bibnamefont{Komatsu}} \bibnamefont{et~al.}
  (\bibinfo{collaboration}{WMAP}), \bibinfo{journal}{Astrophys. J. Suppl.}
  \textbf{\bibinfo{volume}{180}}, \bibinfo{pages}{330} (\bibinfo{year}{2009}),
  \eprint{0803.0547}.

\bibitem[{\citenamefont{Ade et~al.}(2015)}]{Ade:2015rim}
\bibinfo{author}{\bibfnamefont{P.~A.~R.} \bibnamefont{Ade}}
  \bibnamefont{et~al.} (\bibinfo{collaboration}{Planck})
  (\bibinfo{year}{2015}), \eprint{1502.01590}.

\bibitem[{\citenamefont{Guzzo et~al.}(2008)}]{Guzzo:2008ac}
\bibinfo{author}{\bibfnamefont{L.}~\bibnamefont{Guzzo}} \bibnamefont{et~al.},
  \bibinfo{journal}{Nature} \textbf{\bibinfo{volume}{451}},
  \bibinfo{pages}{541} (\bibinfo{year}{2008}), \eprint{0802.1944}.

\bibitem[{\citenamefont{Weinberg}(1989)}]{Weinberg:1988cp}
\bibinfo{author}{\bibfnamefont{S.}~\bibnamefont{Weinberg}},
  \bibinfo{journal}{Rev. Mod. Phys.} \textbf{\bibinfo{volume}{61}},
  \bibinfo{pages}{1} (\bibinfo{year}{1989}).

\bibitem[{\citenamefont{Copeland et~al.}(2006)\citenamefont{Copeland, Sami, and
  Tsujikawa}}]{Copeland:2006wr}
\bibinfo{author}{\bibfnamefont{E.~J.} \bibnamefont{Copeland}},
  \bibinfo{author}{\bibfnamefont{M.}~\bibnamefont{Sami}}, \bibnamefont{and}
  \bibinfo{author}{\bibfnamefont{S.}~\bibnamefont{Tsujikawa}},
  \bibinfo{journal}{Int. J. Mod. Phys.} \textbf{\bibinfo{volume}{D15}},
  \bibinfo{pages}{1753} (\bibinfo{year}{2006}), \eprint{hep-th/0603057}.

\bibitem[{\citenamefont{Ratra and Peebles}(1988)}]{Ratra:1987rm}
\bibinfo{author}{\bibfnamefont{B.}~\bibnamefont{Ratra}} \bibnamefont{and}
  \bibinfo{author}{\bibfnamefont{P.~J.~E.} \bibnamefont{Peebles}},
  \bibinfo{journal}{Phys. Rev.} \textbf{\bibinfo{volume}{D37}},
  \bibinfo{pages}{3406} (\bibinfo{year}{1988}).

\bibitem[{\citenamefont{Wetterich}(1988)}]{Wetterich:1987fm}
\bibinfo{author}{\bibfnamefont{C.}~\bibnamefont{Wetterich}},
  \bibinfo{journal}{Nucl. Phys.} \textbf{\bibinfo{volume}{B302}},
  \bibinfo{pages}{668} (\bibinfo{year}{1988}).

\bibitem[{\citenamefont{Alimi et~al.}(2012)}]{Deus2012}
\bibinfo{author}{\bibfnamefont{J.-M.} \bibnamefont{Alimi}}
  \bibnamefont{et~al.}, in \emph{\bibinfo{booktitle}{High Performance
  Computing, Networking, Storage and Analysis (SC), 2012 International
  Conference for}} (\bibinfo{year}{2012}), pp. \bibinfo{pages}{1--11}, ISSN
  \bibinfo{issn}{2167-4329}.

\bibitem[{\citenamefont{Bouillot et~al.}(2015)}]{Bouillot2015}
\bibinfo{author}{\bibfnamefont{V.~R.} \bibnamefont{Bouillot}}
  \bibnamefont{et~al.}, \bibinfo{journal}{Mon. Not. Roy. Astron. Soc.}
  \textbf{\bibinfo{volume}{450}}, \bibinfo{pages}{145} (\bibinfo{year}{2015}),
  \eprint{1405.6679}.

\bibitem[{\citenamefont{{Springel} et~al.}(2005)}]{2005Nature}
\bibinfo{author}{\bibfnamefont{V.}~\bibnamefont{{Springel}}}
  \bibnamefont{et~al.}, \bibinfo{journal}{Nature}
  \textbf{\bibinfo{volume}{435}}, \bibinfo{pages}{629} (\bibinfo{year}{2005}),
  \eprint{astro-ph/0504097}.

\bibitem[{\citenamefont{Wang and Steinhardt}(1998)}]{Wang1998}
\bibinfo{author}{\bibfnamefont{L.}~\bibnamefont{Wang}} \bibnamefont{and}
  \bibinfo{author}{\bibfnamefont{P.~J.} \bibnamefont{Steinhardt}},
  \textbf{\bibinfo{volume}{1}}, \bibinfo{pages}{19} (\bibinfo{year}{1998}),
  ISSN \bibinfo{issn}{0004-637X}, \eprint{9804015},
  \urlprefix\url{http://arxiv.org/abs/astro-ph/9804015}.

\bibitem[{\citenamefont{Voit}(2005)}]{Voit2004}
\bibinfo{author}{\bibfnamefont{G.~M.} \bibnamefont{Voit}},
  \bibinfo{journal}{Rev. Mod. Phys.} \textbf{\bibinfo{volume}{77}},
  \bibinfo{pages}{207} (\bibinfo{year}{2005}), \eprint{astro-ph/0410173}.

\bibitem[{\citenamefont{Horellou and Berge}(2005)}]{Horellou2005}
\bibinfo{author}{\bibfnamefont{C.}~\bibnamefont{Horellou}} \bibnamefont{and}
  \bibinfo{author}{\bibfnamefont{J.}~\bibnamefont{Berge}},
  \bibinfo{journal}{Monthly Notices of the Royal Astronomical Society}
  \textbf{\bibinfo{volume}{360}}, \bibinfo{pages}{1393} (\bibinfo{year}{2005}),
  ISSN \bibinfo{issn}{00358711}, \eprint{0504465},
  \urlprefix\url{http://arxiv.org/abs/astro-ph/0504465}.

\bibitem[{\citenamefont{Abramo et~al.}(2009)\citenamefont{Abramo, Batista,
  Liberato, and Rosenfeld}}]{Abramo2009}
\bibinfo{author}{\bibfnamefont{L.~R.} \bibnamefont{Abramo}},
  \bibinfo{author}{\bibfnamefont{R.~C.} \bibnamefont{Batista}},
  \bibinfo{author}{\bibfnamefont{L.}~\bibnamefont{Liberato}}, \bibnamefont{and}
  \bibinfo{author}{\bibfnamefont{R.}~\bibnamefont{Rosenfeld}},
  \bibinfo{journal}{Physical Review D} \textbf{\bibinfo{volume}{023516}},
  \bibinfo{pages}{1} (\bibinfo{year}{2009}),
  \urlprefix\url{http://prd.aps.org/abstract/PRD/v79/i2/e023516}.

\bibitem[{\citenamefont{Fernandes et~al.}(2012)\citenamefont{Fernandes,
  de~Carvalho, Kamenshchik, Moschella, and da~Silva}}]{Fernandes2012}
\bibinfo{author}{\bibfnamefont{R.~a.~a.} \bibnamefont{Fernandes}},
  \bibinfo{author}{\bibfnamefont{J.~P.~M.} \bibnamefont{de~Carvalho}},
  \bibinfo{author}{\bibfnamefont{a.~Y.} \bibnamefont{Kamenshchik}},
  \bibinfo{author}{\bibfnamefont{U.}~\bibnamefont{Moschella}},
  \bibnamefont{and} \bibinfo{author}{\bibfnamefont{a.}~\bibnamefont{da~Silva}},
  \bibinfo{journal}{Physical Review D} \textbf{\bibinfo{volume}{85}},
  \bibinfo{pages}{083501} (\bibinfo{year}{2012}), ISSN
  \bibinfo{issn}{1550-7998},
  \urlprefix\url{http://link.aps.org/doi/10.1103/PhysRevD.85.083501}.

\bibitem[{\citenamefont{Li and Xu}(2014)}]{Li2014}
\bibinfo{author}{\bibfnamefont{W.}~\bibnamefont{Li}} \bibnamefont{and}
  \bibinfo{author}{\bibfnamefont{L.}~\bibnamefont{Xu}}, \bibinfo{journal}{Eur.
  Phys. J.} \textbf{\bibinfo{volume}{C74}}, \bibinfo{pages}{2870}
  (\bibinfo{year}{2014}), \eprint{1401.1270}.

\bibitem[{\citenamefont{Mota and van~de Bruck}(2004)}]{Mota2004}
\bibinfo{author}{\bibfnamefont{D.}~\bibnamefont{Mota}} \bibnamefont{and}
  \bibinfo{author}{\bibfnamefont{C.}~\bibnamefont{van~de Bruck}},
  \bibinfo{journal}{Astron.Astrophys.} \textbf{\bibinfo{volume}{421}},
  \bibinfo{pages}{71} (\bibinfo{year}{2004}), \eprint{astro-ph/0401504}.

\bibitem[{\citenamefont{Nunes and Mota}(2006)}]{Nunes2004}
\bibinfo{author}{\bibfnamefont{N.~J.} \bibnamefont{Nunes}} \bibnamefont{and}
  \bibinfo{author}{\bibfnamefont{D.~F.} \bibnamefont{Mota}},
  \bibinfo{journal}{Mon. Not. Roy. Astron. Soc.}
  \textbf{\bibinfo{volume}{368}}, \bibinfo{pages}{751} (\bibinfo{year}{2006}),
  \eprint{astro-ph/0409481}.

\bibitem[{\citenamefont{Wintergerst and Pettorino}(2010)}]{Wintergerst2010}
\bibinfo{author}{\bibfnamefont{N.}~\bibnamefont{Wintergerst}} \bibnamefont{and}
  \bibinfo{author}{\bibfnamefont{V.}~\bibnamefont{Pettorino}},
  \bibinfo{journal}{Phys. Rev.} \textbf{\bibinfo{volume}{D82}},
  \bibinfo{pages}{103516} (\bibinfo{year}{2010}), \eprint{1005.1278}.

\bibitem[{\citenamefont{Bertschinger and Zukin}(2008)}]{Bertschinger:2008zb}
\bibinfo{author}{\bibfnamefont{E.}~\bibnamefont{Bertschinger}}
  \bibnamefont{and} \bibinfo{author}{\bibfnamefont{P.}~\bibnamefont{Zukin}},
  \bibinfo{journal}{Phys. Rev.} \textbf{\bibinfo{volume}{D78}},
  \bibinfo{pages}{024015} (\bibinfo{year}{2008}), \eprint{0801.2431}.

\bibitem[{\citenamefont{Chang et~al.}(2014)\citenamefont{Chang, Lu, and
  Xu}}]{Chang:2014bea}
\bibinfo{author}{\bibfnamefont{B.}~\bibnamefont{Chang}},
  \bibinfo{author}{\bibfnamefont{J.}~\bibnamefont{Lu}}, \bibnamefont{and}
  \bibinfo{author}{\bibfnamefont{L.}~\bibnamefont{Xu}}, \bibinfo{journal}{Phys.
  Rev.} \textbf{\bibinfo{volume}{D90}}, \bibinfo{pages}{103528}
  (\bibinfo{year}{2014}).

\bibitem[{\citenamefont{Shibata and Sasaki}(1999)}]{Shibata1999}
\bibinfo{author}{\bibfnamefont{M.}~\bibnamefont{Shibata}} \bibnamefont{and}
  \bibinfo{author}{\bibfnamefont{M.}~\bibnamefont{Sasaki}},
  \bibinfo{journal}{Phys. Rev. D} \textbf{\bibinfo{volume}{60}},
  \bibinfo{pages}{084002} (\bibinfo{year}{1999}),
  \urlprefix\url{http://link.aps.org/doi/10.1103/PhysRevD.60.084002}.

\bibitem[{\citenamefont{Torres et~al.}(2014)\citenamefont{Torres, Alcubierre,
  Diez-Tejedor, and N\'u\~nez}}]{Torres2014}
\bibinfo{author}{\bibfnamefont{J.~M.} \bibnamefont{Torres}},
  \bibinfo{author}{\bibfnamefont{M.}~\bibnamefont{Alcubierre}},
  \bibinfo{author}{\bibfnamefont{A.}~\bibnamefont{Diez-Tejedor}},
  \bibnamefont{and}
  \bibinfo{author}{\bibfnamefont{D.}~\bibnamefont{N\'u\~nez}},
  \bibinfo{journal}{Phys. Rev. D} \textbf{\bibinfo{volume}{90}},
  \bibinfo{pages}{123002} (\bibinfo{year}{2014}),
  \urlprefix\url{http://link.aps.org/doi/10.1103/PhysRevD.90.123002}.

\bibitem[{\citenamefont{Alcubierre et~al.}(2015)\citenamefont{Alcubierre, de~la
  Macorra, Diez-Tejedor, and Torres}}]{Alcubierre:2015ipa}
\bibinfo{author}{\bibfnamefont{M.}~\bibnamefont{Alcubierre}},
  \bibinfo{author}{\bibfnamefont{A.}~\bibnamefont{de~la Macorra}},
  \bibinfo{author}{\bibfnamefont{A.}~\bibnamefont{Diez-Tejedor}},
  \bibnamefont{and} \bibinfo{author}{\bibfnamefont{J.~M.}
  \bibnamefont{Torres}}, \bibinfo{journal}{Phys. Rev.}
  \textbf{\bibinfo{volume}{D92}}, \bibinfo{pages}{063508}
  (\bibinfo{year}{2015}), \eprint{1501.06918}.

\bibitem[{\citenamefont{Padmanabhan}(1993)}]{padmanabhan1993structure}
\bibinfo{author}{\bibfnamefont{T.}~\bibnamefont{Padmanabhan}},
  \emph{\bibinfo{title}{Structure Formation in the Universe}}
  (\bibinfo{publisher}{Cambridge University Press}, \bibinfo{year}{1993}), ISBN
  \bibinfo{isbn}{9780521424868},
  \urlprefix\url{http://books.google.be/books?id=44gA8634YrEC}.

\bibitem[{\citenamefont{Weinberg and Kamionkowski}(2003)}]{Weinberg2003}
\bibinfo{author}{\bibfnamefont{N.}~\bibnamefont{Weinberg}} \bibnamefont{and}
  \bibinfo{author}{\bibfnamefont{M.}~\bibnamefont{Kamionkowski}},
  \bibinfo{journal}{Monthly Notices of the \ldots}  (\bibinfo{year}{2003}),
  \urlprefix\url{http://mnras.oxfordjournals.org/content/341/1/251.short}.

\bibitem[{\citenamefont{Caldwell et~al.}(1997)\citenamefont{Caldwell, Dave, and
  Steinhardt}}]{Caldwell1997}
\bibinfo{author}{\bibfnamefont{R.~R.} \bibnamefont{Caldwell}},
  \bibinfo{author}{\bibfnamefont{R.}~\bibnamefont{Dave}}, \bibnamefont{and}
  \bibinfo{author}{\bibfnamefont{P.~J.} \bibnamefont{Steinhardt}},
  p.~\bibinfo{pages}{4} (\bibinfo{year}{1997}), ISSN \bibinfo{issn}{0031-9007},
  \eprint{9708069}, \urlprefix\url{http://arxiv.org/abs/astro-ph/9708069}.

\bibitem[{\citenamefont{Hwang and Noh}(2001)}]{Hwang2001}
\bibinfo{author}{\bibfnamefont{J.-c.} \bibnamefont{Hwang}} \bibnamefont{and}
  \bibinfo{author}{\bibfnamefont{H.}~\bibnamefont{Noh}},
  \bibinfo{journal}{Phys.Rev.} \textbf{\bibinfo{volume}{D64}},
  \bibinfo{pages}{103509} (\bibinfo{year}{2001}), \eprint{astro-ph/0108197}.

\bibitem[{\citenamefont{Lasky and Lun}(2007)}]{Lasky2006_2}
\bibinfo{author}{\bibfnamefont{P.~D.} \bibnamefont{Lasky}} \bibnamefont{and}
  \bibinfo{author}{\bibfnamefont{A.~W.} \bibnamefont{Lun}},
  \bibinfo{journal}{Phys.Rev.} \textbf{\bibinfo{volume}{D75}},
  \bibinfo{pages}{024031} (\bibinfo{year}{2007}), \eprint{gr-qc/0612007}.

\bibitem[{\citenamefont{Cordero-Carri\'{o}n and
  Cerd\'{a}-Dur\'{a}n}(2012)}]{Cordero-Carrion2012}
\bibinfo{author}{\bibfnamefont{I.}~\bibnamefont{Cordero-Carri\'{o}n}}
  \bibnamefont{and}
  \bibinfo{author}{\bibfnamefont{P.}~\bibnamefont{Cerd\'{a}-Dur\'{a}n}},
  p.~\bibinfo{pages}{25} (\bibinfo{year}{2012}), \eprint{arXiv:1211.5930},
  \urlprefix\url{http://arxiv.org/abs/1211.5930}.

\bibitem[{\citenamefont{Cordero-Carri\'on and
  Cerd\'a-Dur\'an}(2014)}]{Cordero2014}
\bibinfo{author}{\bibfnamefont{I.}~\bibnamefont{Cordero-Carri\'on}}
  \bibnamefont{and}
  \bibinfo{author}{\bibfnamefont{P.}~\bibnamefont{Cerd\'a-Dur\'an}}, in
  \emph{\bibinfo{booktitle}{Advances in Differential Equations and
  Applications}}, edited by
  \bibinfo{editor}{\bibfnamefont{F.}~\bibnamefont{Casas}} \bibnamefont{and}
  \bibinfo{editor}{\bibfnamefont{V.}~\bibnamefont{Mart\'inez}}
  (\bibinfo{publisher}{Springer International Publishing},
  \bibinfo{year}{2014}), vol.~\bibinfo{volume}{4} of
  \emph{\bibinfo{series}{SEMA SIMAI Springer Series}}, pp.
  \bibinfo{pages}{267--278},
  \urlprefix\url{http://dx.doi.org/10.1007/978-3-319-06953-1_26}.

\bibitem[{\citenamefont{Frieman et~al.}(1995)\citenamefont{Frieman, Hill,
  Stebbins, and Waga}}]{Frieman:1995pm}
\bibinfo{author}{\bibfnamefont{J.~A.} \bibnamefont{Frieman}},
  \bibinfo{author}{\bibfnamefont{C.~T.} \bibnamefont{Hill}},
  \bibinfo{author}{\bibfnamefont{A.}~\bibnamefont{Stebbins}}, \bibnamefont{and}
  \bibinfo{author}{\bibfnamefont{I.}~\bibnamefont{Waga}},
  \bibinfo{journal}{Phys. Rev. Lett.} \textbf{\bibinfo{volume}{75}},
  \bibinfo{pages}{2077} (\bibinfo{year}{1995}), \eprint{astro-ph/9505060}.

\bibitem[{\citenamefont{Amendola and Tsujikawa}(2010)}]{amendola2010}
\bibinfo{author}{\bibfnamefont{L.}~\bibnamefont{Amendola}} \bibnamefont{and}
  \bibinfo{author}{\bibfnamefont{S.}~\bibnamefont{Tsujikawa}},
  \emph{\bibinfo{title}{Dark Energy: Theory and Observations}}
  (\bibinfo{publisher}{Cambridge University Press}, \bibinfo{year}{2010}), ISBN
  \bibinfo{isbn}{9781139488570}.

\end{thebibliography}

\end{document}